\documentclass[aps,pra,twocolumn,amsmath,amssymb,footinbib,showpacs,longbibliography,superscriptaddress]{revtex4-1}

\newcommand{\Jnature}{Nature (London)}
\newcommand{\Jnatphys}{Nat. Phys.}

\newcommand{\Jscience}{Science}

\newcommand{\Jprl}{Phys. Rev. Lett.}

\newcommand{\Jpra}{Phys. Rev. A}
\newcommand{\Jprb}{Phys. Rev. B}

\newcommand{\Jpre}{Phys. Rev. E}

\newcommand{\Jepl}{Europhys. Lett.}
\newcommand{\Jnjp}{New J. Phys.}

\newcommand{\Jepjst}{Eur. Phys. J. Special Topics}

\newcommand{\JApplOpt}{Appl. Opt.}

\newcommand{\JZhEkspTeorFiz}{Zh. Eksp. Teor. Fiz.}

\newcommand{\Jphysrep}{Phys. Rep.}
\newcommand{\JRepProgPhys}{Rep. Prog. Phys.}

\newcommand{\JjphysA}{J. Phys. A: Math. Theor.}

\newcommand{\Jphystoday}{Phys. Today}

\usepackage[english]{babel}
\usepackage{latexsym}
\usepackage{graphics}
\usepackage{subfigure}
\usepackage{epsfig}
\usepackage{color}
\usepackage{hyperref}

\usepackage{braket} 

\hypersetup{
colorlinks=true,
citecolor=blue,
linkcolor=red,
urlcolor=black
}

\newcommand{\subfigimg}[3][,]{%
  \setbox1=\hbox{\includegraphics[#1]{#3}}
  \leavevmode\rlap{\usebox1}
  \rlap{\hspace*{-1pt}\raisebox{\dimexpr\ht1-0.5\baselineskip}{#2}}
  \phantom{\usebox1}
}

\newcommand{\ie}{{i.e.}}
\newcommand{\eg}{{e.g.}}

\renewcommand{\Re}{\operatorname{Re}}
\renewcommand{\Im}{\operatorname{Im}}

\newcommand{\Ur}{U_{\textrm{\tiny R}}}
\newcommand{\sigmar}{\sigma_{\textrm{\tiny R}}}

\DeclareMathOperator{\dd}{\mathrm{d}\!}

\DeclareMathOperator{\e}{e}

\renewcommand{\subsubsection}[1]{\textit{#1.---}}

\begin{document}

\title{Expansion of a quantum wave packet in a one-dimensional disordered potential in the presence of a uniform bias force}

\author{C.~Crosnier~de~Bellaistre}
\affiliation{
 Centre de Physique Th\'eorique, Ecole Polytechnique, CNRS, Univ Paris-Saclay, F-91128 Palaiseau, France
}
\affiliation{
 Laboratoire Charles Fabry,
 Institut d'Optique, CNRS, Univ Paris-Saclay,
 2 avenue Augustin Fresnel,
 F-91127 Palaiseau cedex, France
}

\author{C.~Trefzger}
\affiliation{
 Laboratoire Charles Fabry,
 Institut d'Optique, CNRS, Univ Paris-Saclay,
 2 avenue Augustin Fresnel,
 F-91127 Palaiseau cedex, France
}

\affiliation{
European Commission, DG Communications Networks, Content and Technology, Directorate C - 'Digital Excellence and Scientific Infrastructures', EUFO 2/256, L-2920 Luxembourg
}

\author{A.~Aspect}
\affiliation{
 Laboratoire Charles Fabry,
 Institut d'Optique, CNRS, Univ Paris-Saclay,
 2 avenue Augustin Fresnel,
 F-91127 Palaiseau cedex, France
}

\author{A.~Georges}
\affiliation{
 Centre de Physique Th\'eorique, Ecole Polytechnique, CNRS, Univ Paris-Saclay, F-91128 Palaiseau, France
 }
\affiliation{Coll\`ege de France, 11 place Marcelin Berthelot, 75005 Paris, France}
\affiliation{Center for Computational Quantum Physics, Flatiron Institute, 162 Fifth avenue, New York, NY 10010, USA}
\affiliation{Department of Quantum Matter Physics, University of Geneva, 24 Quai Ernest-Ansermet, 1211 Geneva 4, Switzerland}

\author{L.~Sanchez-Palencia}
\affiliation{
 Centre de Physique Th\'eorique, Ecole Polytechnique, CNRS, Univ Paris-Saclay, F-91128 Palaiseau, France
}

\date{\today}

\begin{abstract}
We study numerically the expansion dynamics of an initially confined quantum wave packet in the presence of a disordered potential and a uniform bias force. For white-noise disorder, we find that the wave packet develops asymmetric algebraic tails for any ratio of the force to the disorder strength. The exponent of the algebraic tails decays smoothly with that ratio and no evidence of a critical behavior on the wave density profile is found. Algebraic localization features a series of critical values of the force-to-disorder strength where the $m$-th position moment of the wave packet diverges. Below the critical value for the $m$-th moment, we find fair agreement between the asymptotic long-time value of the $m$-th moment and the predictions of diagrammatic calculations. Above it, we find that the $m$-th moment grows algebraically in time. For correlated disorder, we find evidence of systematic delocalization, irrespective to the model of disorder. More precisely, we find a two-step dynamics, where both the center-of-mass position and the width of the wave packet show transient localization, similar to the white-noise case, at short time and delocalization at sufficiently long time. This correlation-induced delocalization is interpreted as due to the decrease of the effective de Broglie wave length, which lowers the effective strength of the disorder in the presence of finite-range correlations.
\end{abstract}


\maketitle

\section{Introduction}
Anderson localization of coherent classical or quantum waves in disordered media is by now a well-established phenomenon. It has recently received strong theoretical and experimental assessment for a variety of systems~\cite{abrahams2010,lagendijk2009,aspect2009,lsp2010,modugno2010}.
In a homogeneous system, it is characterized by exponential suppression of transmission and absence of diffusion or expansion. Both take place on a unique length scale, known as the localization length. The latter is essentially determined by the strength of the disorder and the energy of the wave.
An immediate consequence of Anderson localization is that the static conductivity, which characterizes the current response to an electric force, vanishes at zero temperature.
However, this result follows from linear-response theory and holds in the limit where the force vanishes.
Less is known about the impact of a finite bias force on localization, but consensus is by now established on two main effects.
On the one hand, localization can survive but is strongly suppressed.
While exponential spatial decay of wave functions in the absence of a force entails strong localization, the presence of a finite force entails a much weaker form of localization where wave functions decay only algebraically, at least in one dimension~\cite{delyon1984}.
On the other hand, localization in the presence of a force lacks complete universality. For instance, the power of the algebraic decay has been shown to significantly differ in transmission and expansion schemes~\cite{prigodin1980,soukoulis1983,perel1984,ccdb2017a}.

Algebraic localization is expected to be the strongest in one-dimensional geometry since the bias force field only couples states that are all localized in the absence of the force~\cite{kirkpatrick1986}. Moreover, analytic calculations are possible in this case~\cite{berezinskii1974,gogolin1976a,gogolin1976b,lifshits1988}
and precise conclusions can be drawn.
For a typical transmission scheme, where a plane wave enters a white-noise disordered medium of finite extension, the transmission coefficient decays algebraically with an exponent proportional to the disorder strength and inversely proportional to the bias force, irrespective to their ratio $\alpha$~\cite{soukoulis1983,perel1984,ccdb2017a}.
For an expansion scheme, where an initially strongly confined wave packet is released into a disordered medium of infinite extension, the density profile acquires asymptotically in time an algebraic spatial decay below some critical value of the ratio of the force to the disorder, $\alpha=1$~\cite{prigodin1980}. Above it, the theory cannot be normalized and it is expected that the wave packet is delocalized.
These predictions follow from the properties of the stationary density profile at infinite time.
In contrast, little is known about the time-dependent dynamics of the wave packet, either towards algebraic localization or towards infinite expansion for weak or strong bias force respectively,
as well as about the critical behavior at the transition point.
Similarly, little is known about the effect of finite-range disorder correlations.

In this paper, we study numerically the expansion dynamics of a wave packet initially confined in an arbitrary small region and released into white-noise and correlated disordered potentials.
For white-noise disorder, we find that the expanding wave packet develops rapidly algebraic tails. For $\alpha<1$, it reaches a stationary profile and we find an exponent of the algebraic decay in good agreement with the analytical results of Ref.~\cite{prigodin1980}. For $\alpha>1$, we still find density profiles with algebraic tails on a finite spatial range. The latter increases in time while the average density decreases continuously.
The exponent of the algebraic tails varies smoothly around the expected transition $\alpha=1$ and shows no sign of a singular behavior.
Nevertheless, algebraic localization entails a series of critical values $\alpha_m$, characterized by the divergence of the $m$-th position moment of the expanding wave packet~\cite{prigodin1980}.
We find that the $m$-th moment shows clear critical behavior at $\alpha_m$ for $m=1$ and $m=2$. They signal transition towards absence of global motion and absence of expansion, respectively.
For correlated disorder, we show that the wave packet is always delocalized. More precisely, we identify a two step dynamics, where the $m$-th moment first shows transient localization similar to the white-noise case and then delocalization. This correlation-induced delocalization effect is attributed to the decrease of the effective de Broglie wave length, which lowers the effective strength of the disorder in the presence of finite-range correlations.

The paper is organized as follows.
In Sec.~\ref{sec:ProbaTrans}, we review the results of Ref.~\cite{prigodin1980}, which will be useful in the following.
Then, we discuss the numerical results on the expansion dynamics
for white-noise and correlated disorder in Secs.~\ref{sec:WhiteNoise} and~\ref{sec:correlated_disorder}, respectively.
We finally summarize our results and discuss possible observation in ultracold-atom experiments such as those of Refs.~\cite{billy2008,roati2008,kondov2011,jendrzejewski2012} in Sec.~\ref{sec:conclusion}.

\section{Probability of transfer in white-noise disorder}
\label{sec:ProbaTrans}
In this section, we set the problem and briefly review the results of Ref.~\cite{prigodin1980} on the asymptotic, infinite-time, algebraic localization of a one-dimensional (1D) quantum wave in a disordered potential. Writing the equation of motion in dimensionless units, we show that the dynamics beyond the characteristic time associated with the wave energy depends on a single parameter, which we identify.

\subsection{Spreading of a dragged wave packet in a disordered potential}
\begin{figure}
\includegraphics[width=0.49\textwidth]{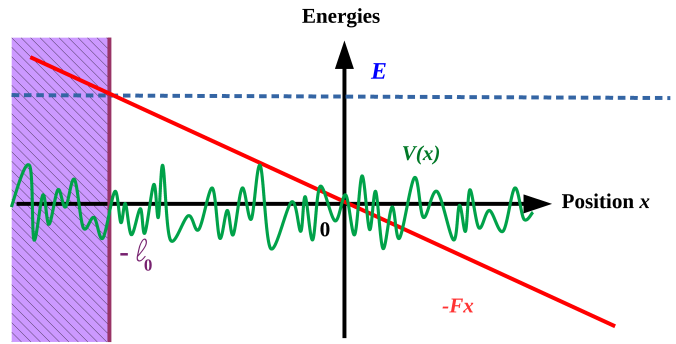}
\caption{(Color online) Scheme of the system.
The quantum wave packet has an energy $E$ (dashed blue line). It propagates in a disordered potential $V(x)$ (solid green line) in the presence of a linear bias potential $-Fx$ (solid red line).
The classically forbidden region, $x<-\ell_0$ is represented by the purple hatched zone.
}
\label{fig:scheme}
\end{figure}

Consider a 1D, non interacting, quantum wave packet subjected to a disordered potential $V(x)$ and a uniform, constant force $F$ (see Fig.~\ref{fig:scheme}).
It is described by the wave function $\psi(x,t)$, governed by the Schr\"odinger equation
$i\hbar\partial_t\psi(x,t) = \hat{H} \psi(x,t)$
with the Hamiltonian
\begin{equation}\label{eq:Hamiltonian}
\hat{H}=\frac{-\hbar^2\nabla^2}{2m} + V(\hat{x}) - F \hat{x},
\end{equation}
where $m$ is the mass of the particle, $x$ is the position, and $t$ is the time.
We assume that the bias force is positive, $F \geq 0$, so that the particle is dragged towards the right. While we restrict ourselves to 1D geometry, the discussion also applies to elongated confining guides, provided the disordered potential $V(x)$ is transversally invariant~\cite{billy2008,roati2008,brantut2012}.
Without loss of generality, we set the energy reference such that the disorder average is null, $\overline{V(x)}=0$, where the overbar denotes disorder averaging.
We assume that the disordered potential is Gaussian and homogeneous (for details on the practical implementation, see Sec.~\ref{sec:WhiteNoise:NumApproach}). It is characterized by the two-point correlation function $C(x)=\overline{V(x')V(x'+x)}$, independent of the reference point $x'$, and higher-order correlation functions are found using the Wick theorem~\cite{lebellac1991}.
For white-noise disorder, we write $C(x)=\Ur\delta(x)$, where $\Ur$ is the disorder strength.
For correlated disorder, we write $C(x)=(\Ur/\sigmar)c(x/\sigmar)$, where $\sigmar$ is the correlation length while the function $c(u)$ is normalized by $\int du\, c(u)=1$  and has a width of order unity.

We consider the expansion of a wave packet initially confined close to the origin $x=0$ at time $t=0$.
The disorder-averaged density profile is $\overline{n}(x,t) = \textrm{Tr}\{ \hat{\rho}(t) \hat{n}(x)\}$, with $\hat{\rho}$ the one-body density matrix and $\hat{n}(x) = \delta(x-\hat{x})$ the spatial density operator.
Within semi-classical approximation, it reads as
\begin{equation}
\overline{n}(x,t) = \int \dd E \int \dd x_0 \ W_0 (x_0,E) P(x,t|x_0,E),
\end{equation}  
where $W_0(x_0,E)$ represents the semi-classical, position-energy joint probability distribution of the initial wave packet
and
\begin{equation}
\label{eq:p_def_correlators}
P(x,t \vert x_0,E) = \frac{\overline{\braket{E|\delta[x-\hat{x}(t)]\delta[x_0-\hat{x}]|E}}}{\overline{\braket{E|\delta(\hat{x})|E} }},
\end{equation}
with $\ket{E}$ the eigenstate of $\hat{H}$ of energy $E$,
represents the probability of transfer  of a particle of energy $E$ from the initial point $x_0$ to the final point $x$ in a time $t$~\cite{skipetrov2008,piraud2011}.
The semi-classical approach has been shown to describe the dynamics of the wave packet with very good accuracy~\cite{lsp2007,*lsp2007erratum,billy2008,jendrzejewski2012,piraud2012a,piraud2013b}.
Hence, knowing the initial wave packet, the dynamics is entirely determined by the energy-resolved probability of transfer $P(x,t \vert x_0,E)$. In the following, we focus on the latter.

\subsection{Probability of transfer}\label{sec:ProbTransf}
We now write the probability of transfer in dimensionless units and identify the minimal parameters relevant to the problem.

\subsubsection{Dimensionless form}
For white-noise disorder, the probability of transfer $P(x,t)$ depends on two variables, namely, the position $x$ and the time $t$, and four parameters, namely, the force $F$, the mass $m$, the disorder strength $\Ur$, and the particle energy $E$~\cite{corrdis}.
To get rid of two of these parameters, it is fruitful to introduce the natural classical length and time scales,
\begin{equation}
\ell_0 \equiv E / F
\qquad , \qquad
t_0 \equiv {\sqrt{2mE}}/{F},
\end{equation}
respectively.
The length $\ell_0$ is the opposite of the disorder-free classical turning point, which is the point where the classical velocity $v$ of a particle of energy $E=mv^2/2-Fx$, without disorder, vanishes (see Fig.~\ref{fig:scheme}).
The time $t_0$ is the time to reach this point from the $x=0$ with the energy $E$ and a left-pointing initial velocity.
Then, using the dimensionless distance and time,
\begin{equation}\label{eq:ellANDtau}
\xi \equiv x/\ell_0
\qquad , \qquad
\tau \equiv t/t_0,
\end{equation}
the rescaled wave function $\phi(\xi,\tau)=\sqrt{\ell_0}\psi(x,t)$,
and the dimensionless parameters
\begin{equation}\label{eq:epsilonANDalpha}
\varepsilon\equiv \hbar F/\sqrt{2m}E^{3/2}
\qquad , \qquad
\alpha\equiv {\hbar^2 F}/{m\Ur},
\end{equation}
the Schr\"odinger equation reduces to
\begin{equation}
i\frac{\partial \phi}{\partial \tau} = \varepsilon \frac{\partial^2\phi}{\partial \xi^2} + v(\xi) \phi - \frac{\xi}{\varepsilon}\phi,
\end{equation}
where $v(\xi) \equiv (t_0/\hbar) V(\xi\ell_0)$ is the disordered potential in dimensionless units.
The parameter $\varepsilon$ stands for both the inverse effective mass and the inverse effective force. The parameter $\alpha$ stands for the inverse effective disorder strength, $\overline{v(\xi)v(\xi')}=(2/\alpha)\delta(\xi-\xi')$.
Hence, the dimensionless probability of transfer reads as
\begin{equation}
p(\xi,\tau|\alpha,\varepsilon) \equiv \ell_0 \times P(x,t|\Ur,m,F,E)
\end{equation}
and only depends on the two dimensionless parameters $\alpha$ and $\varepsilon$.

\subsubsection{Asymptotic long-time behavior}
\label{sec:asymptbehavior}
Further examination indicates that one can get rid of one more parameter in the long-time limit.
The probability of transfer of the particle in white-noise disorder may be calculated using diagrammatic expansion~\cite{prigodin1980}. In the presence of a bias force, one finds that the solution $p(\xi,\tau|\alpha,\varepsilon)$ depends on $\varepsilon$ only via the quantity
\begin{eqnarray}
\nu(s,\omega \vert \alpha,\varepsilon)
& = &
\frac{4\alpha}{\varepsilon}(1+\xi)^{3/2}\left[ \left(1 + \frac{\omega \varepsilon}{1+\xi}  \right)^{3/2} - 1 \right.
\nonumber \\
& & \left. - \frac{\omega \varepsilon}{1+\xi} \left( 1 + \frac{\omega \varepsilon}{1+\xi} \right)^{1/2} \right].
\nonumber
\end{eqnarray}
Then, in the long-time limit, $\omega \ll 1/\varepsilon$, one finds
\begin{equation}
\nonumber
\nu(s,\omega|\alpha,\varepsilon) \simeq 2\alpha\omega \sqrt{1+\xi}.
\end{equation}
Hence, the dependence on the parameter $\varepsilon$ disappears.
This result, confirmed by the numerical simulations presented below (see Sec.~\ref{sec:density_profile}), extends the analytical result at infinite time~\cite{prigodin1980}.

\subsection{Localization solution}
\label{sec:prigodin_sol}
We finally review previous results for the infinite-time limit, which will be useful in the remainder of the paper.

\subsubsection{General solution}
The probability of transfer has been solved analytically in the infinite-time limit in Ref~\cite{prigodin1980}.

For a moderate force, $\alpha<1$, and in the classically allowed region, $\xi \geq -1$, one finds
\begin{equation}
\label{eq:prigodin_exact}
p^\pm_{\infty}(\xi|\alpha)=\frac{\pi\sin(\pi\alpha)}{32\alpha^2(1\!+\!\xi)^{\eta_\pm}}
\int_0^{\infty}\!\dd\lambda\;f(\alpha,\lambda)(1+\xi)^{\mp\frac{\lambda^2}{8\alpha}}
\end{equation}
with 
\begin{equation}
\label{eq:falpha}
f(\alpha,\lambda)\equiv\lambda\sinh(\pi\lambda)\frac{(1\!+\!\alpha^2\!+\!\lambda^2)^2\!-\!4\alpha^2}{[\cosh(\pi\lambda)\!+\!\cos(\pi\alpha)]^2}
\end{equation}
and 
\begin{equation}
\label{eq:eta_pm}
\eta_\pm \equiv 1 \pm \frac{(1 \mp \alpha)^2}{8\alpha},
\end{equation}
where the upper signs holds for $\xi \geq 0$ and the lower sign for $-1<\xi<0$.
Note that, for any value of $\alpha$, the probability distribution~(\ref{eq:prigodin_exact}) is normalized, $\int \dd \xi\, p(\xi\vert\alpha)=1$.
In the direction opposite to the force, the probability of transfer is disregarded beyond the classical turning point, $\xi < -1$, since it vanishes exponentially. Note that Eq.~\eqref{eq:prigodin_exact} gives $p_\infty^-(\xi=-1)=0$ for $\alpha<3-2\sqrt{2} \simeq 0.17$ but $p(\xi)$ diverges when $\xi \rightarrow -1$ for $\alpha>3-2\sqrt{2}$.

Note that Eq.~(\ref{eq:prigodin_exact}) generalizes the exact result of Ref.~\cite{berezinskii1974} for the probability of transfer in 1D non-biased disorder to the case where a bias force is present.
One can also check that the vanishing force limit of Eq.~(\ref{eq:prigodin_exact}) matches the result of Ref.~\cite{berezinskii1974} and shows exponential localization (see Appendix~\ref{sec:Andersonlimit}).

For $\alpha \geq 1$, the probability of transfer cannot be normalized (see below) and, to our knowledge, no analytical form is known.

\subsubsection{Algebraic localization and critical exponents}
In the infinite distance limit, $\xi \to \infty$, the probability of transfer decays algebraically, $p_\infty^+(\xi|\alpha) \sim 1/\xi^{\eta_+}$, up to logarithmic corrections~\cite{prigodin1980}. 
While in the absence of a bias force, Anderson localization is characterized by the exponential decay of the probability of transfer and, consequently, all the position moments are finite, in the presence of a bias force, one finds a weaker form of localization where only the lowest-order position moments,
\begin{equation}
\label{eq:xi_m_prigo}
\overline{\braket{\xi^m}}_{\infty}=\int_{-1}^\infty \dd\xi\, p_\infty(\xi|\alpha)\xi^m,
 \end{equation} 
can be finite, owing to the algebraic decay of the probability of transfer.
More precisely, the $m$-th moment is finite only for $\eta_+ \ge m+1$~\cite{mthmomentfinte}.
It defines a series of critical values of the parameter $\alpha$,
\begin{equation}
\alpha_m=1+4m-2\sqrt{4m^2+2m},
\end{equation}
such that $\overline{\braket{\xi^m}}_{\infty}$ diverges for $\alpha>\alpha_m$, \ie\ when the ratio of the force to the disorder strength exceeds an $m$-dependent critical value. When $\alpha \geq 1$, all the position moments diverge. It includes the $m=0$ moment, $\overline{\braket{\xi^0}}_{\infty}=\int_{-1}^\infty \dd\xi\, p_\infty(\xi|\alpha)$, which is the normalization of the probability of transfer.
This points towards a full delocalization of the particle  when $\alpha$ becomes larger than 1.
Table~\ref{tab:alphac} shows the values of the first critical parameters $\alpha_m$ for several values of $m$.

\begin{table}
\begin{tabular}{p{0.75cm} | p{0.1cm} p{0.5cm} p{0.95cm} p{0.95cm} p{0.95cm} p{0.95cm} p{0.95cm} p{0.25cm}}
\hline \hline
$m$         && 0     & 1		& 2         & 3		& 4       & $\infty$ \\
\hline
$\alpha_m$  && 1     & 0.101     & 0.056    & 0.039    & 0.029  & 0 \\
\hline \hline
\end{tabular}
\caption{Values of the critical parameters $\alpha_m$ of divergence of the moments of the position $\braket{\xi^m}_\infty$.}
\label{tab:alphac}
\end{table}

\section{Spreading of a wave packet in white-noise disorder}
\label{sec:WhiteNoise}
In this section, we study the expansion dynamics of the quantum wave packet,  initially confined in a very small region of space, in the presence of white-noise disorder and of a uniform bias force.
We first describe the numerical approach. We then study the time evolutions of the center-of-mass position and of the width of the wave packet.

\subsection{Numerical approach}
\label{sec:WhiteNoise:NumApproach}
	
\subsubsection{Initial wave packet}
We consider the expansion dynamics of the initial Gaussian wave packet
\begin{equation}
\phi_0(\xi) = \e^{i\kappa_0\xi}\frac{\e^{-\xi^2/2\Delta\xi^2}}{\pi^{1/4}\sqrt{\Delta\xi}}.
\end{equation}
It is centered at the dimensionless position $\xi=0$
with the positive momentum $\kappa_0=1/\varepsilon$.
The real-space and momentum widths are, respectively, $\Delta \xi$ and $\Delta \kappa = 1/\Delta\xi$.
In the following, we assume that the initial wave packet is well localized in space and energy so that we can assimilate the dimensionless density profile, $\rho(\xi,\tau) \equiv \ell_0 n(x,t)$, to the probability of transfer,
\begin{equation}\label{eq:condA}
\rho(\xi,\tau) \simeq p(\xi,\tau \vert \alpha, \varepsilon).
\end{equation}
It requires that the real-space and energy widths are small enough.
On the one hand, since the localization is algebraic, there is no typical length scale for the probability of transfer. Hence, the influence of the initial spatial width on the density profile will be negligible at distances exceeding it,
\begin{equation}
\xi \gg \Delta \xi.
\end{equation}
On the other hand, the energy width, $\Delta E$, is controlled by the initial momentum width, $\Delta E = \hbar^2 k_0 \Delta k / m$ with $k_0=\kappa_0/\ell_0$, and the disorder-induced spectral broadening, $\Delta E_\textrm{\tiny dis} = \hbar/t_-$, where $t_-=\sqrt{2\hbar^4E/m\Ur^2}$ is the mean free scattering time. Our assumptions require that the typical energy $E \simeq \hbar^2 k_0^2/2m$ exceeds both. In dimensionless units, the conditions read as
\begin{equation}\label{eq:condB}
\Delta \kappa \ll \kappa_0.
\end{equation}
and 
\begin{equation}\label{eq:condC}
\varepsilon \ll \alpha.
\end{equation}
Note that the condition~(\ref{eq:condC}) ensures that the probability of transfer is independent of the parameter $\varepsilon$ beyond the propagation time $\tau \simeq \alpha$ at most (see Sec.~\ref{sec:ProbTransf}).
In practice, we use the values $k/\kappa_0=0.1$ and $\varepsilon/\alpha=0.1$, except whenever explicitly mentioned. Then, the initial width of the wave packets has negligible influence for, at most, $\xi \gg \alpha$.

\subsubsection{Time evolution}
To study the dynamical evolution of the wave packet, we use exact numerical diagonalization of the Hamiltonian~(\ref{eq:Hamiltonian}) for a given realization of the disordered potential (see below). We determine the eigenenergies $E_j$ (in units of $\hbar t_0$) and the associated dimensionless eigenfunctions $\Phi_j(\xi)$, where $j$ spans the spectrum.
We then project the initial wave packet $\phi_0(\xi)$ onto the eigenstates $\Phi_j(\xi)$ and write
\begin{equation}\label{eq:decomp}
\phi(\xi,\tau)=\sum_{j}\Phi_j(\xi) \e^{-i E_j \tau} \braket{\Phi_j|\phi_0}.
\end{equation}
The overlap of the initial state and the energy eigenstates is significant in a limited part of the spectrum. In practice, we use a cut-off on the energy of the eigenvectors and restrict the sum in Eq.~(\ref{eq:decomp}) to the eigenstates with  energy $-E \lesssim E_j \leq 2E$ or $0 \lesssim E_j \leq 5E$ depending on the value of $\alpha$~\cite{lapack}.
We have checked that in all the cases considered below, the overlap of the initial wave packet with the chosen eigenstates exceeds $99\%$.

The system length $L$ is chosen so as to match the classical position reached at $t_{\max}$, \ie\ 
\begin{equation}
x_{\max}=Ft_{\max}^2/2m+\hbar k_0 t_{\max}/m.
\end{equation}
The 1D space is discretized with a length unit $\delta x$, which satisfies $\delta x \le 0.1 \lambda(x_{\max})$, where $\lambda(x)=2\pi\hbar/\sqrt{2m(E+F x)}$ is the wave length associated to the classical kinetic energy of the particle at position $x$.
The number of points on the space grid thus scales as
$N \sim x_{\max} / \delta x \sim \xi_{\max}^{3/2}/\varepsilon$ for large distances, $\xi_{\max} \gg 1$.
In order to reach the highest value of $\xi_{\max}$ with a given maximal number of spatial grid points $N$, one should use the highest value of $\varepsilon$. To satisfy also the weak disorder condition~\eqref{eq:condC}, we use $\varepsilon/\alpha = 0.1$ in all the numerics presented below, except whenever explicitly mentioned.
In some cases, we use $\varepsilon/\alpha=0.01$ and confirm that the time evolution of the density profile is independent of $\varepsilon/\alpha$ for large enough times.
In order to satisfy the narrow momentum width condition~(\ref{eq:condB}), we use $\Delta\kappa/\kappa_0 \simeq 0.14$.

\subsubsection{Disorder}
To produce an homogeneous Gaussian disorder with null statistical average, $\overline{V(x)}=0$, and the two-point correlation function $C(x)=\overline{V(x')V(x'+x)}$ numerically, we use standard techniques (see for instance Refs.~\cite{huntley1989,horak1998,cheng2002}). We first generate a complex random field $g(k)$. The real $\Re[g(k)]$ and imaginary $\Im[g(k)]$ parts of it are independent Gaussian random variables. They
satisfy $\Re[g(-k)]=\Re[g(k)]$, $\overline{\Re[g(k)]}=0$, and $\overline{\Re[g(k)]\Re[g(k')]}=\frac{1}{2}\tilde{C}(k)\delta(k-k')+\frac{1}{2}\tilde{C}(k)\delta(k+k')$ for the real part, and
$\Im[g(-k)]=-\Im[g(k)]$, $\overline{\Im[g(k)]}=0$, and $\overline{\Im[g(k)]\Im[g(k')]}=\frac{1}{2}\tilde{C}(k)\delta(k-k')-\frac{1}{2}\tilde{C}(k)\delta(k+k')$ for the imaginary part, where $\tilde{C}(k)$ is the Fourier transform of $C(x)$. 
Numerically, these conditions are easily satisfied on the discrete reciprocal space $[-j \times\delta k, j\times\delta k]$ by generating for all $j>0$ two random Gaussian variables $\Re[g(j\times\delta k)]$ and $\Im[g(j\times\delta k)]$  of null average and of variance equal to $\frac{1}{2}\tilde{C}(k)$, and then taking $\Re[g(-j\times\delta k)]=\Re[g(j\times\delta k)]$ and $\Im[g(-j\times\delta k)]=-\Im[g(j\times\delta k)]$. 
 The disorder is finally obtained by Fourier transform of the field $g(k)$. Note that for a white-noise disorder, $\tilde{C}(k)=\Ur$ is a constant.
  
The final results are then averaged over hundreds of realizations of the disorder.

\subsection{Evolution of the density profile}
\label{sec:density_profile}
\begin{figure}
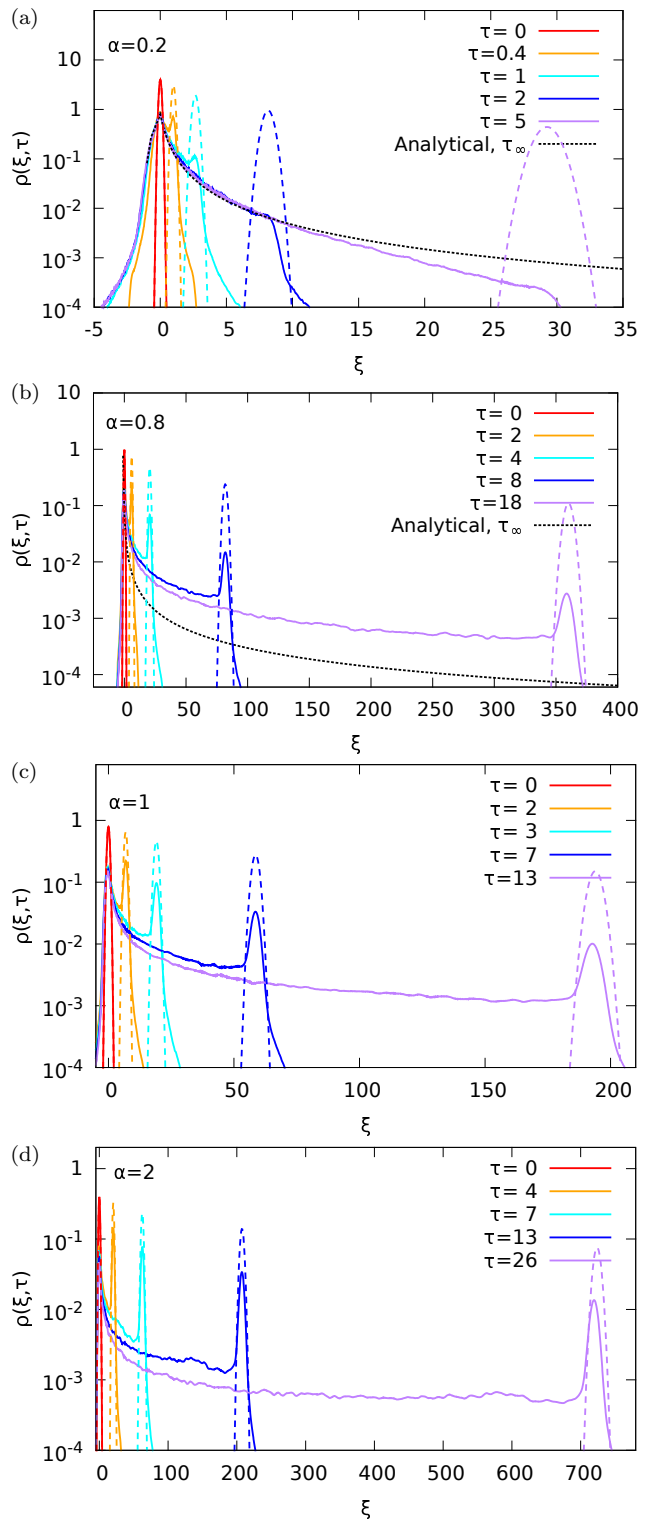

  \begin{tabular}{@{}p{0.96\linewidth}@{}}
    \subfigimg[width=\linewidth]{\footnotesize{\color{black} (a)}}{fig2a_profile_a_0_2} \\
    \subfigimg[width=\linewidth]{\footnotesize{\color{black} (b)}}{fig2b_profile_a_0_8} \\
    \subfigimg[width=\linewidth]{\footnotesize{\color{black} (c)}}{fig2c_profile_a_1} \\
    \subfigimg[width=\linewidth]{\footnotesize{\color{black} (d)}}{fig2d_profile_a_2} \\
  \end{tabular}
\caption{(Color online)
Time evolution of the density profile of an initially Gaussian quantum wave packet expanding in the presence of white-noise disorder and a uniform bias force.
The various panels correspond to different ratios of the force to the disorder: (a)~$\alpha=0.2$, (b)~$\alpha=0.8$, (c)~$\alpha=1$, and (d)~$\alpha=2$.
The solid color lines correspond to different times indicated on each panel, with the peak propagating towards the right when time increases.
The asymptotic probability of transfer (dotted black lines for $\alpha<1$) and the time evolution of the wave packet subjected to the force and in the absence of disorder (dashed color lines) are also shown.
}
\label{fig:density}
\end{figure}

In Fig.~\ref{fig:density}, we plot the density $\rho(\xi,\tau)$ in lin-log scale as a function of the position $\xi$ at different times $\tau$.
The various panels correspond to different values of the ratio of the force to the disorder strength, $\alpha$ (increasing from top to bottom).
The asymptotic probability of transfer~\eqref{eq:prigodin_exact} (dotted black lines) and the spreading of the same wave packet dragged by the force in the absence of disorder (dashed color lines) are also shown for reference.
Note that the former exists only for $\alpha<1$ (upper two panels in Fig.~\ref{fig:density}).
The latter is found analytically:
\begin{equation}
\rho(\xi,\tau)=|\phi(\xi,\tau)|^2=\frac{1}{\sqrt{2\pi}\sigma_{\xi,0}(\tau)}\e^{-[x-{\xi_{0}(\tau)}]^2/2\sigma_{\xi,0}^2(\tau)}
\end{equation}
with
\begin{equation}
{\xi_{0}(\tau)} = \tau^2+2\tau, \label{eq:xi0}
\end{equation}
and 
\begin{equation}
\sigma_{\xi,0}(\tau)
=\frac{\varepsilon}{\sqrt{2}}\frac{\kappa_0}{\Delta\kappa}\sqrt{1+4\frac{\tau^2}{\varepsilon^2}\left(\frac{\Delta\kappa}{\kappa_0}\right)^4},\label{eq:sigma0}
\end{equation}
(see Appendix~\ref{sec:appendixB}).

The initial wave packet is extremely narrow and centered around the origin, $\xi=0$.
It slightly expands in the direction opposite to the drag force, $\xi<0$, and rapidly reaches a stationary density profile in that direction. The latter approximately matches the theoretical asymptotic probability of transfer $p_\infty(\xi \vert \alpha)$ [see Eq.~\eqref{eq:prigodin_exact}] in the classically allowed region, $\xi > -1$.

The wave packet expands mostly in the direction of the bias force, $\xi > 0$, where the behavior is richer.
For small values of $\alpha$ [Fig.~\ref{fig:density}(a)], the density profile of the expanding wave packet smoothly approaches the asymptotic profile. For short time and short distance, it is close to the latter. At longer distance, it shows a sharp edge located around the position of the disorder-free expanded profile (dashed color lines). At longer times, the edge gets smoother and eventually disappears. Then, the profile at long distance progressively reaches the asymptotic profile from below at the expense of the density at short distance (hardly visible on the vertical logarithmic scale).

For larger values of $\alpha$ [Figs.~\ref{fig:density}(b)-\ref{fig:density}(d)], the edge of the propagating wave packet remains sharp on very long times. Moreover, it is marked by a clear density peak that reproduces the shape of the wave packet expanding under the bias force in the absence of disorder, although with a reduced amplitude. It is easily interpreted as the fraction of the wave packet that has not yet been scattered with the disordered potential. The amplitude of the edge density peak decreases with time since a larger fraction of the wave packet interacts with the disorder and is back-scattered as found in the numerics. The leakage of the front density peak progressively constructs a profile at shorter distance that reaches the asymptotic profile when it exists, \ie\ for $\alpha<1$ [Fig.~\ref{fig:density}(b)]. In contrast to the behavior observed for smaller values of $\alpha$, the latter is here reached from above.

Similar dynamics is observed for values of $\alpha$ exceeding the critical value $\alpha_0=1$, where the asymptotic probability of transfer~(\ref{eq:prigodin_exact}) is no longer normalized  [Figs.~\ref{fig:density}(c) and \ref{fig:density}(d)]. In this case, however, the density develops an almost flat profile in the long-distance limit, truncated around the position of disorder-free expanded wave packet.
Its amplitude decreases when the edge propagates so as required by the conservation of the particle number.

In order to study the evolution of the wave packet more precisely, it is fruitful to define the dynamically rescaled density profile
\begin{equation}
\tilde{\rho}(u,\tau)= \xi_0(\tau) \rho \big(u\xi_0(\tau),\tau\big),
\end{equation}
where $\xi_0(\tau)=\tau^2+2\tau$ is the central position of the disorder-free expanding wave packet [see Eq.~\eqref{eq:xi0}]. Except for very small values of $\alpha$, this dynamically-rescaled density profile shows a sharp edge at $u \simeq 1$, almost independent of time.

\begin{figure}
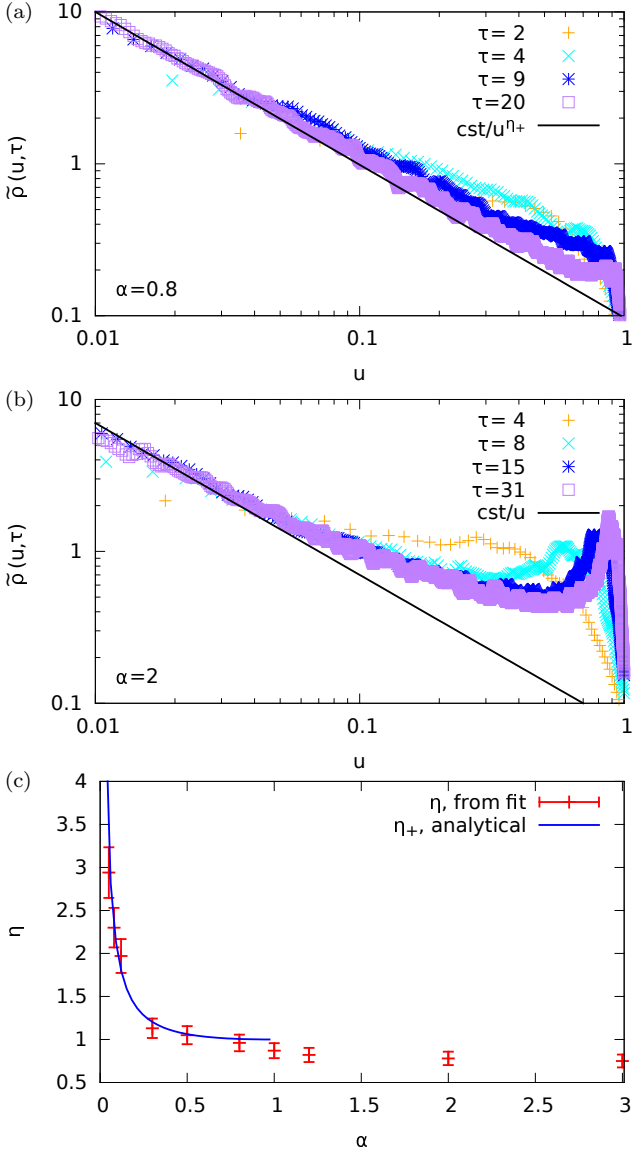

  \begin{tabular}{@{}p{0.96\linewidth}@{}}
    \subfigimg[width=\linewidth]{\footnotesize{\color{black} (a)}}{fig3a_profile-rescale} \\
    \subfigimg[width=\linewidth]{\footnotesize{\color{black} (b)}}{fig3b_profile-rescale}
    \subfigimg[width=\linewidth]{\footnotesize{\color{black} (c)}}{fig3c_fit-eta}
  \end{tabular}
\caption{(Color online)
Panels (a) and (b) show the dynamically rescaled density $\tilde{\rho}$ as a function of the dynamically rescaled position $u$ at different times $\tau$ indicated on each panel.
They are compared to the algebraic decays of slope $1/u^{\eta_+}$ for $\alpha<1$ and $1/u$ for $\alpha > 1$ (solid black lines).
Panel~(a) corresponds to $\alpha=0.8$ and panel~(b) to $\alpha=2$. In the numerics, we used negative initial velocities, $\kappa_0<0$.
(c)~Values of $\eta$ obtained by a linear fit of the dynamically-renormalized density in log-log scale on $u\in [0.01,0.1]$, with 10\% error bars.
The solid blue line shows the analytical prediction $\eta_+$ for $0 < \alpha < \alpha_0=1$ [Eq.~(\ref{eq:eta_pm})].}
\label{fig:profil_superpose_log_log}
\end{figure}

Figure~\ref{fig:profil_superpose_log_log} shows, in a log-log scale, the dynamically rescaled density profile $\tilde{\rho}(u,\tau)$ as a function of the rescaled position $u$ at different times $\tau$ and for two values of $\alpha$.
Note that in order to reduce the amplitude of the edge peak around $u=1$, we used here a negative initial velocity, $\kappa_0<0$, in the numerics. This way, only the small part of the wave packet that has reached the classical turning point can move towards the right without scattering on the disordered potential.
The dynamically rescaled profiles are compared to the algebraic decay $1/u^{\eta}$,
with $\eta=\eta_+$ for $\alpha<1$ [Fig.~\ref{fig:profil_superpose_log_log}(a), $\alpha=0.8$] and  $\eta=\lim_{\alpha \rightarrow 1_-}\eta_+=1$ for $\alpha>1$ [Fig.~\ref{fig:profil_superpose_log_log}(b), $\alpha=2$].
For $\alpha<1$, we find that the dynamically rescaled density profile agrees with the predicted power-law decay already at short times.
The profiles collapse on the same line at short distance.
Deviations are observed at distances approaching the profile edge, $u \simeq 1$, which, however, vanish in the long-time limit.
For $\alpha>1$, we also observe data collapse at short distance. In this case, the dynamically rescaled density profile shows a decay that is weaker than $1/u$.

In Fig.~\ref{fig:profil_superpose_log_log}(c), we plot the exponent $\eta$ found by fitting the power law $\tilde{\rho}(u,\tau) = A(\tau)/u^\eta$ to the density profile found in the numerics. We find that the exponent $\eta$ decreases smoothly as a function of $\alpha$. The fitted values (red points) are in good agreement with the analytical prediction~(\ref{eq:eta_pm}) (see solid blue line) in the whole validity range of the theory, $0<\alpha<1$. In this regime where the asymptotic probability of transfer is normalizable, we find that the fitted amplitude $A(\tau)$ reaches a constant value.
For $\alpha<1$,  we still find an algebraic decay with, however, $\eta<1$ so that the amplitude $A(\tau)$ decays in time. We find no sign of any critical behavior of the exponent $\eta$ at the critical point $\alpha_0=1$.

\subsection{Evolution of the position moments}
\begin{figure*}
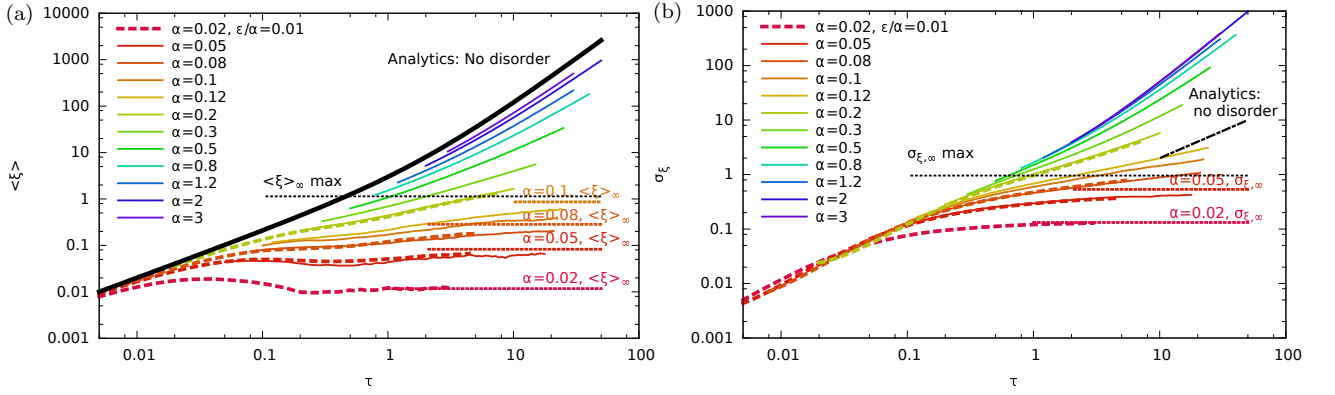

  \begin{tabular}{@{}p{0.48\linewidth}@{}p{0.48\linewidth}@{}}
    \subfigimg[width=\linewidth]{\footnotesize{\color{black} (a)}}{fig4a_x} &
    \subfigimg[width=\linewidth]{\footnotesize{\color{black} (b)}}{fig4b_sigmax}
  \end{tabular}
\caption{\label{fig:evolcumulants}(Color online)
Time evolution of (a)~the center of mass position and (b)~width of the wave packet for various values of the parameter $\alpha$ and, for some curves, two values of $\varepsilon$ (solid lines correspond to $\varepsilon/\alpha=0.1$ and dashed lines to $\varepsilon/\alpha=0.01$).
The lower (upper) curves correspond to the smallest (largest) values of $\alpha$ indicated on the panels.
Also shown are the analytical prediction in the absence of disorder (black lines), the asymptotic values $\overline{\xi}_{\infty}$ and  $\sigma_{\xi,\infty}$ (horizontal solid colorful segments) and their maximum values (horizontal dotted black lines).
}
\end{figure*}

The expansion dynamics of the wave packet may be further characterized by the time evolution of its position moments,
\begin{equation}
\overline{\xi^m}(\tau)=\int \dd \xi \, \rho (\xi,\tau) \xi^m,
\end{equation}
with $m \in \mathbb{N}$,
and the corresponding cumulants.
Figure~\ref{fig:evolcumulants} shows the numerical solution for the evolution of the first two cumulants $\overline{\xi}$~[Fig.~\ref{fig:evolcumulants}(a)] and $\sigma_\xi=\sqrt{\overline{\xi^2}-\overline{\xi}^2}$~[Fig.~\ref{fig:evolcumulants}(b)] as a function of the dimensionless time $\tau$ for various values of the parameter $\alpha$.
Comparison of the time evolutions for $\varepsilon/\alpha=0.1$ (solid lines) and  $\varepsilon/\alpha=0.01$ (dashed lines) confirm that the dynamics is nearly independent of the parameter $\varepsilon$ (for $\tau \gg \varepsilon$, see Sec.~\ref{sec:asymptbehavior}).
Both $\overline{\xi}$ and $\sigma_\xi$ essentially increase with time, which indicates that the wave packet advances in the direction of the force and spreads. As expected, the increase is faster for larger values of $\alpha$, that is, when the force $F$ increases with respect to the disorder strength $\Ur$. We now need to distinguish the two cases $\alpha < \alpha_m$ and $\alpha > \alpha_m$ for both center of mass ($\overline{\xi}$; $m=1$) and width ($\sigma_\xi$; $m=2$).

\subsubsection{Localized regime ($\alpha < \alpha_m$)}
\begin{figure}
\includegraphics[width=0.49\textwidth]{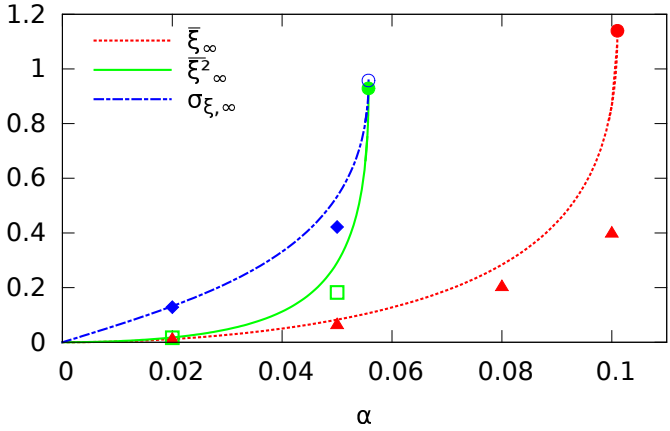}
\caption{(Color online)
Values at large time of the first two position moments $\overline{\xi}$ and $\overline{\xi^2}$ and the width $\sigma_{\xi}$ of the position of the wave packet, as a function of $\alpha$. The lines represent the analytical values at infinite time and the maximum values of $\overline{\xi}_\infty$ at $\alpha=\alpha_1$ and $\overline{\xi^2}_\infty$ and $\sigma_{\xi,\infty}$ at $\alpha=\alpha_2$ are shown as circles. The diamonds, squares and triangles correspond respectively to the values of $\overline{\xi}(\tau_{\max})$, $\overline{\xi^2}(\tau_{\max})$ and $\sigma_\xi(\tau_{\max})$ at the highest time $\tau_{\max}$ calculated in the numerical simulations.
}
\label{fig:x_x2_sigma_theoric}
\end{figure}

In the localized regime for the position moment $m$, \ie\ $\alpha<\alpha_m$,
$\overline{\xi^m}$ is finite. It can be calculated analytically using Eq.~\eqref{eq:xi_m_prigo}, together with Eqs.~(\ref{eq:prigodin_exact})-(\ref{eq:eta_pm}). It yields
\begin{equation}
\overline{\xi}_{\infty} = 24\pi\alpha\sin(\pi\alpha) g(\alpha,8)
\end{equation}
and
\begin{equation}
\overline{\xi^2}_{\infty} = 16\pi\alpha\sin(\pi\alpha) \left[ 5 g(\alpha,16)-3 g(\alpha,8) \right]
\end{equation}
with
\begin{eqnarray}
g(\alpha,c) & \equiv & \int_0^{\infty} \dd \lambda\ \frac{\sinh(\pi\lambda)}{[\cosh(\pi\lambda)+\cos(\pi\alpha)]^2}
\label{eq:functg} \\
&&  \times \left(\frac{\lambda}{\lambda^2+(1-\alpha)^2-c\alpha}+\frac{\lambda}{\lambda^2+(1+\alpha)^2+c\alpha}\right).
\nonumber
\end{eqnarray}
The two quantities $\overline{\xi}_{\infty}$ and $\overline{\xi^2}_{\infty}$ are continuous, increasing functions of $\alpha\in[0,\alpha_m]$, and reach a finite value at $\alpha=\alpha_m$, see Fig.~\ref{fig:x_x2_sigma_theoric}.
For instance, we find $\max \left(\overline{\xi}_{\infty}\right) \simeq 1.140$, $\max \left(\overline{\xi^2}_{\infty}\right) \simeq  0.929$, and $\max \left(\sigma_{\xi,\infty}\right) \simeq 0.958$.
Then, they both diverge for $\alpha>\alpha_m$.
Note that all the position moments $\overline{\xi^m}$ vanish in the limit $\alpha=0$.
It is expected in the case where $\alpha$ vanishes as a result of a diverging disorder strength $\Ur$ for a fixed force $F$ and a fixed energy $E$. Then the particle is infinitely localized at its initial position.
In contrast, the case where $\alpha$ vanishes due to a vanishing force for a fixed energy and a fixed disorder strength requires more care.
Then, we expect the average position to remain null but all other position moments should be finite due to a finite exponential localization length.
This is actually consistent with our results since the moments of the position are given by $\overline{\braket{x^m}}_{\infty}=E^m\overline{\braket{\xi^m}}_{\infty}/F^m$, and they thus correspond to a finite value when $F=0$.
For instance, for $\alpha \rightarrow 0$,
we find
$\overline{\xi}_{\infty} \simeq 24\pi^2g_0\alpha^2$,
\ie\ $\overline{x}_{\infty} \rightarrow 0$,
and $\overline{\xi^2}_{\infty} \simeq 32\pi^2 g_0 \alpha^2$,
\ie\ $\overline{x^2}_{\infty} \simeq 32\pi^2 g_0 (\hbar^2 E/m\Ur)^2$,
with $g_0 \equiv \lim_{\alpha \rightarrow 0} g(\alpha,c) \simeq 0.122$.

In the numerical calculations shown in Fig.~\ref{fig:evolcumulants}(a),
we find that, after some oscillations, the average position $\overline{\xi}(\tau)$ converges to a constant value for $\alpha \lesssim\alpha_1 \simeq 0.101$.
The values of $\overline{\xi}(\tau_{\max})$ at the larger time that we have calculated, $\tau_{\max}$, are plotted as red triangles on Fig.~\ref{fig:x_x2_sigma_theoric}. They agree well with the infinite-time theoretical value for smallest $\alpha$ and reproduce its trend of increase for larger $\alpha$~\cite{converge}.
As shown in Fig.~\ref{fig:evolcumulants}(b), a similar behavior is found for the width of the wave packet, $\sigma_\xi(\tau)$, for $\alpha \lesssim \alpha_2 \simeq 0.056$, without, however, significant oscillations at intermediate times.
Note that for $\alpha=0.08 \in ]\alpha_2,\alpha_1]$,
the width $\sigma_\xi$ diverges while the center-of-mass position $\overline{\xi}$ converges to a finite value.

\subsubsection{Delocalized regime ($\alpha > \alpha_m$)}
Consider now the delocalized regime for each cumulant, $\alpha > \alpha_m$, and let us focus first on the center of mass position ($\overline{\xi}$; $m=1$). As shown in Fig.~\ref{fig:evolcumulants}(a), in this regime, $\overline{\xi}(\tau)$ no longer saturates to a finite value in the long-time limit.
For $\alpha$ slightly above the critical value $\alpha_1$, the curve for $\overline{\xi}(\tau)$ in log-log scale is nearly linear, which suggests the power law behavior, $\overline{\xi}(\tau) \sim \tau^{\beta_1}$, with some exponent $\beta_1$ that depends on $\alpha$. For values of $\alpha$ significantly above $\alpha_1$, $\overline{\xi}(\tau)$ seems to increase faster than linearly on the considered expansion times plotted on Fig.~\ref{fig:evolcumulants}(a). However, when $\alpha$ increases, that is when the relative strength of the force to the disorder increases, the curves for $\overline{\xi}(\tau)$ approach the analytic solution in the absence of disorder, namely, $\overline{\xi}_0(\tau) = \tau^2+2\tau$ [solid black line in Fig.~\ref{fig:evolcumulants}(a)]. The latter is a power law with the exponent $\beta_1^\infty=2$ in the infinite-time limit. It suggests that in the presence of disorder, $\overline{\xi}(\tau)$ should increase as a power law at most with $\beta_1 \leq 2$, \ie\ the curves in Fig.~\ref{fig:evolcumulants}(a) are asymptotically straight lines in log-log scale.

\begin{figure*}
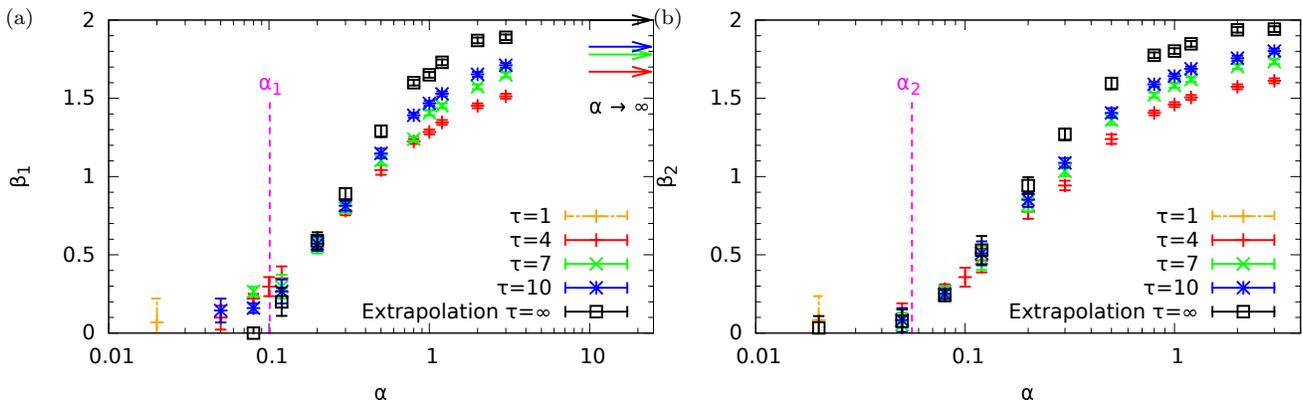

  \begin{tabular}{@{}p{0.48\linewidth}@{}p{0.48\linewidth}@{}}
    \subfigimg[width=\linewidth]{\footnotesize{\color{black} (a)}}{fig6a_beta1} &
    \subfigimg[width=\linewidth]{\footnotesize{\color{black} (b)}}{fig6b_beta2}
  \end{tabular}
\caption{\label{fig:graph_beta_alpha}(Color online)
Power-law exponents of
(a)~the average position and the (b)~the width
of the wave packet as a function of the parameter $\alpha$, at different expansion times $\tau$ (color points) and extrapolated to $\tau\to\infty$ (black points).
The expected values for $\alpha\to\infty$ are signaled by right-pointing arrows in panel~(a).
The error bars for the data at finite time $\tau$ (color points) correspond to the standard deviation due to disorder averaging. The data at infinite time (black points) are extrapolations of the latter as discussed in Appendix~\ref{sec:appendixA}, which produces an additional contribution to the error bars.
}
\end{figure*}

To check it, we define the instantaneous power-law exponent
\begin{equation}
\beta_1 (\tau) \equiv \frac{\dd\, \ln \overline{\xi}(\tau)}{\dd\, \ln \tau}.
\end{equation}
The exponent $\beta_1$ is plotted as a function of $\alpha$ on Fig.~\ref{fig:graph_beta_alpha}(a) at different times $\tau$ (color points) together with their values extrapolated at infinite time (black points, see below). 
For $\alpha \lesssim \alpha_1$, the exponent $\beta_1$ is close to zero and show weak fluctuations versus the time $\tau$.
For $\alpha \gtrsim \alpha_1$ conversely, $\beta_1$ increases with $\alpha$.
These results are compatible with the expected delocalization transition at $\alpha=\alpha_1$.

In the delocalized regime, $\alpha>\alpha_1$, the instantaneous power-law exponent $\beta_1(\tau)$ shows a clear systematic increase with time, which, however, slows down. It suggests that the dynamics of $\overline{\xi}(\tau)$ converges slowly towards a power law in the long-time limit.
This is consistent with the corresponding behavior in the absence of disorder,
\begin{equation}
\beta_1^\infty (\tau) = 2 \frac{\tau+1}{\tau+2},
\end{equation}
which converges only algebraically, $\beta_1^\infty (\tau) \simeq 2 \left(1-1/\tau \right)$, towards the asymptotic value $\beta_1^\infty=2$.
The values of $\beta_1^\infty(\tau)$ are shown as arrows pointing towards $\alpha=\infty$ in Fig.~\ref{fig:graph_beta_alpha}(a). Note that for all times $\tau$ the values of $\beta_1(\tau)$ in the presence of disorder are all smaller than $\beta_1^\infty(\tau)$ and tend to them when $\alpha$ increases.
In order to determine the asymptotic value of the power-law exponent $\beta_1(\infty)$ in the presence of disorder, we plot $\beta_1(\tau)$ as a function of $\beta_1^{\infty}(\tau)$, and use a linear extrapolation at $\beta_1^{\infty}=2$, which corresponds to infinite time $\tau=\infty$ (see appendix~\ref{sec:appendixA}). 
The values of the power-law exponent $\beta_1$ extrapolated at infinite time are shown as black squares on Fig.~\ref{fig:graph_beta_alpha}(a). They confirm localization, \ie\ $\beta_1 \simeq 0$, for $\alpha \lesssim \alpha_1$.
For $\alpha \gtrsim \alpha_1$, the exponent $\beta_1^\infty$ shows a sharp increase right above the transition and converges towards the finite value $\beta_1^\infty=2$ corresponding to the disorder-free case when the relative strength of the force to the disorder increases.

We now focus on the width of the wave packet ($\sigma_\xi^2$; $m=2$). As shown in Fig.~\ref{fig:evolcumulants}(b), a similar behavior as for the average position $\overline{\xi}(\tau)$ is found for the width $\sigma_\xi(\tau)$ for $\alpha > \alpha_2 \simeq 0.056$. We therefore similarly look for a power law behavior, $\sigma_\xi(\tau) \sim \tau^{\beta_2}$, with some exponent $\beta_2$ that depends on $\alpha$. Note that in the absence  of disorder the analytic solution is given by Eq.~\eqref{eq:sigma0}. This result indicates that the width in the absence of disorder depends on the value of $\varepsilon$ at short times. It, however, disappears in the long-time limit, where we find
\begin{equation}
\sigma_{\xi}\simeq \sqrt{2}\frac{\Delta\kappa}{\kappa_0}\tau,
\end{equation}
which corresponds to a power law $\beta_2^\infty=1$ in the infinite-time limit. This asymptotic behavior is plotted with a dotted black line on Fig.~\ref{fig:evolcumulants}(b). We find that contrary to the curves of the average position, the plotted curves for the width do not converge towards the asymptotic solution in the absence of disorder when $\alpha$ increases. It indicates that the disorder remains relevant even for a very strong force.
Moreover, at a given time $\tau$, the width decreases when $\alpha$ increases for large values of $\alpha$ and a given value of $\varepsilon$. On Fig.~\ref{fig:graph_beta_alpha}(b), we plot the values of $\beta_2$ as a function of $\alpha$ at different times $\tau$ (colored dots)  found from the curves of Fig.~\ref{fig:evolcumulants} (b). The extrapolated values at infinite time, found using the same method as for the average position, are shown as black squares. As for the exponent $\beta_1$, the power-law exponent $\beta_2$ shows a sharp transition between the localized region $\alpha\le \alpha_2$, where its numerical values are approximately equal to zero, and the delocalized region $\alpha>\alpha_2$, where it starts to increase with $\alpha$. For $\alpha \gtrsim 0.3$, the value of $\beta_2$ exceeds 1 and approaches 2 for $\alpha \rightarrow \infty$. It is thus larger the infinite-time value in the absence of disorder.
This is confirmed by a further study for very large values of $\alpha$, $\alpha\sim 10^2-10^4$, which shows that for a given $\varepsilon$, at short times, the width increases as in the absence of disorder, and at large times it starts to increase faster and becomes independent of the value of $\varepsilon$. The larger the value of $\alpha$, the later this change of behavior (see Appendix~\ref{sec:appendixA}).

\begin{figure}
\includegraphics[width=0.49\textwidth]{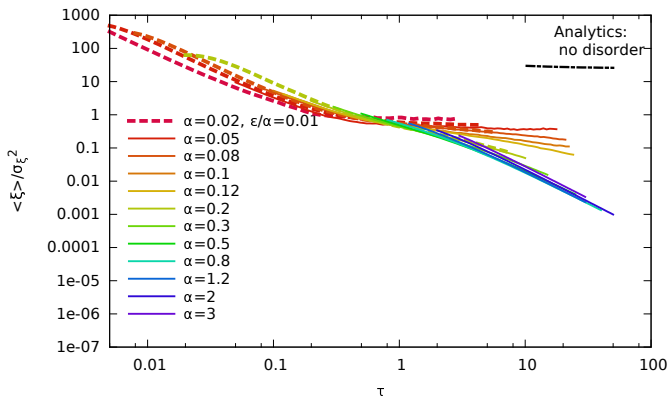}
\caption{(Color online)
Ratio of the center-of-mass position to the width of the wave packet as a function of time for various values of the parameter $\alpha$ and, for some curves, two values of $\varepsilon$ (solid lines correspond to $\varepsilon/\alpha=0.1$ and dashed lines to $\varepsilon/\alpha=0.01$).
The lower (upper) curves on the right-hand side correspond to the largest (smallest) values of $\alpha$ indicated on the panels.
}
\label{fig:einstein}
\end{figure}

Let us finally discuss our results in light of the Einstein's relation. For classical Brownian motion, Einstein's theory relates the linear drift in the presence of a force 
to the spread of the diffusing packet in the absence of a bias, namely: 
$\langle \xi \rangle_\textrm{\tiny cl} / (\sigma_\xi^\textrm{\tiny cl,0} )^2 = \textrm{const}$, 
implying in particular $\beta_1=2\beta_2=1$. 
Even when diffusion is anomalous (non-Brownian, $\beta_2\neq 0.5$), 
it is expected from linear response theory applied to a classical diffusion process that 
this relation holds in the short-time regime~\cite{bouchaud1990}. 
It is thus tempting to explore the validity of this relation for the diffusion of the 
quantum wave packet studied here. To this end, we plot on 
Fig.~\ref{fig:einstein} the ratio $\langle \xi \rangle / \sigma_\xi^2$ as a function of 
time $\tau$. It is seen that the quantity $\langle \xi \rangle / \sigma_\xi^2$ continuously decreases when $\tau$ increases, signaling a breakdown of the Einstein relation at long times for all values of $\alpha$. However, a plateau regime where $\langle \xi \rangle / \sigma_\xi^2 \simeq \textrm{const}$ at intermediate times is indeed observed close to the localized regime $\alpha \sim \alpha_1$. 
This breakdown of the Einstein relation is not surprising since there is no quantum diffusion regime in one dimension and transient diffusion is always non-Brownian. It is indeed known even for classical non-Brownian diffusion processes that Einstein relation can be violated at long times in some cases, \eg in the presence of a wide distribution of trapping times~\cite{bouchaud1990}. 
In contrast, in higher dimension, we would expect for the problem considered here that 
a regime of validity of the Einstein relation holds in the diffusive regime.

\section{Wave packet evolution in correlated disorder}
\label{sec:correlated_disorder} 
 We now turn to a correlated model of disorder.
In the absence of a force, finite-range correlations of the disorder do not alter significantly exponential Anderson localization.
In fact, the effect of finite correlations can be fully encapsulated in the renormalization of the effective disorder strength and, consequently, of the localization length at a given energy or wavelength~\cite{lifshits1988,gogolin1976a,gogolin1976b}.
In the presence of a bias force, however, such a renormalization cannot be applied because the wavelength decreases owing to the dragging by the force.
As shown rigorously in transmission schemes, correlations may induce delocalization for any model disorder~\cite{ccdb2017a}.
Here, we study numerically the effect of a correlated disorder on the expansion of a quantum wave packet.

\subsection{Numerical results}
We still consider a Gaussian, homogeneous disorder and with a null average, but we now assume that the two-point correlation function is a Gaussian function,
\begin{equation}
C(x)=\frac{\Ur}{\sqrt{2\pi}\sigmar}\exp\left(-\frac{x^2}{2\sigmar^2}\right),
\end{equation}
where $\sigmar$ is the correlation length of the disorder. Using the units introduced in Sec.~\ref{sec:ProbTransf}, we define the dimensionless correlation length $\tilde{\sigma}_{\textrm{\tiny R}}=\sigmar/\ell_0$.
In addition to the two dimensionless parameters $\alpha$ and $\varepsilon$, the probability of transfer now depends on a third dimensionless parameter, namely, $\kappa_0\tilde{\sigma}_{\textrm{\tiny R}}$.
In the absence of a force, the localization length takes the form $L_\textrm{\tiny loc} = 4\hbar^2 E / m\tilde{C}(2k_E)$, where $k_E=\sqrt{2mE}/\hbar$ is the wave vector associated to the energy $E$.
For white-noise disorder, the localization length reduces to $L_\textrm{\tiny loc} = 4\hbar^2 E / m\Ur$.
Hence, in the absence of a force, the effect of finite-range correlations amounts to renormalizing the disorder strength $\Ur$ to the effective energy-dependent value $\tilde{C}(2k_E)$.
By analogy, in the presence of a bias force, it is then convenient to quantify the ratio of the force to the disorder strength by the local parameter
\begin{equation}
\alpha_{\mathrm{loc}}(x)=\frac{\hbar^2F}{m\tilde{C}[2k(x)]}.
\end{equation}
Note that one finds
$\alpha_{\mathrm{loc}}(0)={\hbar^2F}/{m\tilde{C}(2k_E)}=\alpha \Ur/\tilde{C}(2k_E)$, hence $\alpha_{\mathrm{loc}}(0)=\alpha \exp(2\kappa_0^2 \tilde{\sigma}^2_{\textrm{\tiny{R}}})$.

We study the evolution of the center-of-mass position $\overline{\xi}(\tau)$ and width $\sigma_\xi(\tau)$ of the wave packet for three values of $\alpha_{\mathrm{loc}}(0)$, corresponding to
(a)~$\alpha_{\mathrm{loc}}(0)<\alpha_2<\alpha_1$,
(b)~$\alpha_2<\alpha_{\mathrm{loc}}(0)<\alpha_1$,
and
(c)~$\alpha_2<\alpha_1<\alpha_{\mathrm{loc}}(0)$,
respectively.
For white-noise disorder with $\alpha=\alpha_{\mathrm{loc}}(0)$, they correspond, respectively, to the cases where
(a)~both $\overline{\xi}(\tau)$ and $\sigma_\xi(\tau)$ are finite,
(b)~$\overline{\xi}(\tau)$ is finite but $\sigma_\xi(\tau)$ diverges,
and (c)~both $\overline{\xi}(\tau)$ and $\sigma_\xi(\tau)$ diverge.
In the numerics, we use $\varepsilon/\alpha=0.1$ and $\kappa_0\tilde{\sigma}_{\textrm{\tiny{R}}} =0.5\alpha$.

\begin{figure*}
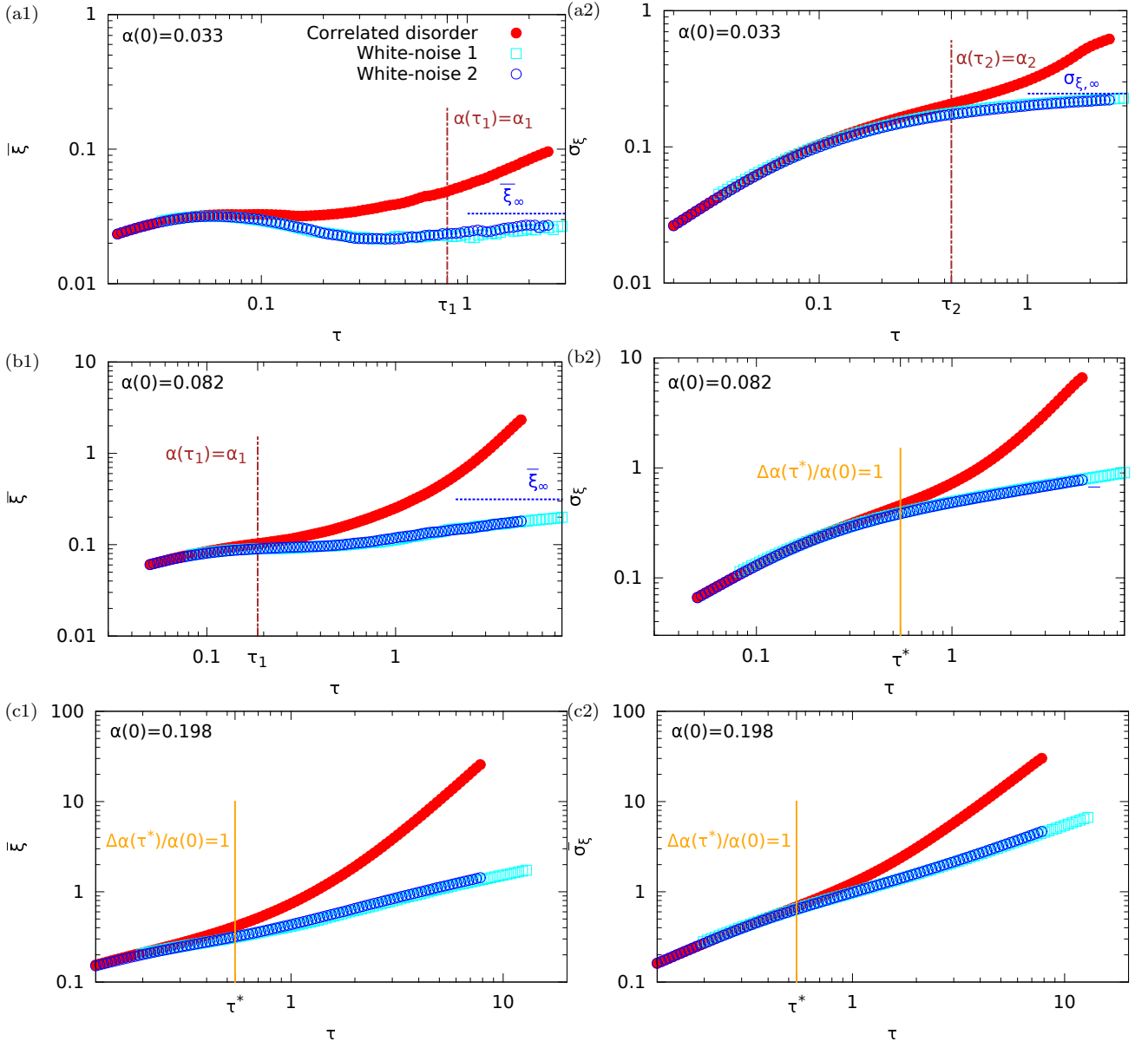

  \begin{tabular}{@{}p{0.48\linewidth}@{}p{0.48\linewidth}@{}}
    \subfigimg[width=\linewidth]{\footnotesize{\color{black} (a1)}}{fig8a1} &
    \subfigimg[width=\linewidth]{\footnotesize{\color{black} (a2)}}{fig8a2}
    \\
    \subfigimg[width=\linewidth]{\footnotesize{\color{black} (b1)}}{fig8b1} &
    \subfigimg[width=\linewidth]{\footnotesize{\color{black} (b2)}}{fig8b2}  
\\
     \subfigimg[width=\linewidth]{\footnotesize{\color{black} (c1)}}{fig8c1} &
    \subfigimg[width=\linewidth]{\footnotesize{\color{black} (c2)}}{fig8c2}     \\
  \end{tabular}
\caption{\label{fig:graph_x_avr_corr}(Color online)
Evolution of the average position $\overline{\xi}(\tau)$ of the wave packet (left panel), and its width $\sigma_\xi(\tau)$ (right panel) as a function of time $\tau$ in the presence of a bias force. The filled red circles correspond to a correlated disorder with the Gaussian two-point correlation function $\tilde{C}(x)=\frac{\Ur}{\sqrt{2\pi}\sigmar}\exp\left(\frac{-x^2}{2\sigmar^2}\right)$, with $\varepsilon/\alpha=0.1$, $\tilde{\sigma}_{\textrm{\tiny{R}}}\kappa_0=0.05\alpha$, and for three different values of $\alpha_{\mathrm{loc}}(0)=\alpha\exp(2\tilde{\sigma}_{\textrm{\tiny{R}}}\kappa_0)$. Empty light blue squares correspond to white-noise disorder with $\alpha=\alpha_{\mathrm{loc}}(0)$ and $\varepsilon/\alpha=0.1$. Filled dark blue circles correspond to white-noise disorder with $\alpha=\alpha_{\mathrm{loc}}(0)$ and $\varepsilon
\exp(2\tilde{\sigma}_{\textrm{\tiny{R}}}^2\kappa_0^2)/\alpha=0.1$. The blue horizontal dashed lines correspond to the asymptotic values of the moments for white-noise disorder when they exist.}
\end{figure*}

The results of the numerical simulations are shown on Fig.~\ref{fig:graph_x_avr_corr} as filled red circles.
Also shown are results of simulations for a white-noise disorder ($\sigmar=0$) with the parameter $\alpha$ set equal to $\alpha_{\mathrm{loc}}(0)$ for comparison.

Numerical results for two values of $\varepsilon$ are presented.
In the first case, we keep the same value of  $\varepsilon/\alpha$ for the expansion in white-noise disorder as for the expansion in correlated disorder. In practice, this is equivalent for a given wave packet (fixed mass and fixed energy) to keep the same disorder strength $\Ur$, but increase the force so that the value of $\alpha$ becomes equal to $\alpha_{\mathrm{loc}}(0)$.  The results are presented as open light blue squares.
In the second case, the value of $\varepsilon$ is the same as for the white-noise disorder. It means that the force is not changed, but the disorder strength $\Ur$ is reduced so as to increase $\alpha$ up to $\alpha_{\mathrm{loc}}(0)$. The results are presented as open dark blue circles.
The results found on the average position (left panel) and the width (right panel) of the wave packet are identical in the white-noise disorder for the two values of $\varepsilon$ tested, as expected at large enough times, but also for short times.

For all the values of $\alpha_{\mathrm{loc}}(0)$ we consider, we find that both $\overline{\xi}$ and $\sigma_{\xi}$ are almost identical for the white-noise disorder and the correlated disorder at short times. Then a tendency to localization is found in the cases where it is expected for white-noise disorder [(a) and (b) for $\overline{\xi}(\tau)$ and only (a) for $\sigma_\xi(\tau)$].
At longer times, however, we find that both $\overline{\xi}(\tau)$ and $\sigma_\xi(\tau)$ start to increase faster in the correlated disorder compared to the white-noise disorder. This indicates that the wave packet crosses over towards delocalization in the correlated disorder, irrespective to its strength with respect to the force.

\subsection{Physical interpretation}  
The correlation-induced delocalization effect can be interpreted using a simple physical picture, inspired by the transmission scheme,
where similar delocalization occurs~\cite{ccdb2017a}. In the transmission scheme, it has been rigorously shown that this delocalization can be understood using a simple semi-classical interpretation. It results from the increase of the semi-classical kinetic energy of the particle, $K(x)=E+Fx$,
and, consequently, of the local mean free path $\ell_-(x)=2\hbar^2 K(x)/ m\tilde{C}[2k(x)]$, where $k(x)=\sqrt{2mK(x)}/\hbar$.

For white-noise disorder, we found that the wave packet approximately expands up to the classical disorder-free position $\xi_0(\tau)=\tau^2+2\tau$ (see Sec.~\ref{sec:density_profile}). Hence, the maximal local value of $\alpha$ felt by the wave packet at time $\tau$ approximately reads as
\begin{equation}
\alpha(\tau)=\alpha_{\mathrm{loc}}[\xi_0(\tau)\ell_0].
\end{equation} 
Note that at the initial time, $\tau=0$, we recover the value $\alpha_{\mathrm{loc}}(0)=\hbar^2F/m\tilde{C}(2k_E)$ introduced previously.
To interpret the behavior of the quantum wave in the correlated disorder, we can then compare the value of $\alpha(\tau)$ to the critical values $\alpha_m$ for the center-of-mass position $\overline{\xi}(\tau)$ ($m=1$) and the width $\sigma_\xi(\tau)$ ($m=2$).
More precisely, we distinguish two situations.

Consider first the cases where the moment $\overline{\xi^m}$ is finite for white-noise disorder, \ie\ $\alpha_{\mathrm{loc}}(0)<\alpha_m$.
In this case, the delocalization of the $m$-th position moment in the correlated disordered potential may be estimated as the time $\tau_m$ where the maximum local value of $\alpha(\tau)$ reaches the delocalization threshold $\alpha_m$, \ie\ $\alpha(\tau_m)=\alpha_m$. We indicate on Fig.~\ref{fig:graph_x_avr_corr} the values of $\tau_1$ [panels~(a1) and (b1)] and $\tau_2$ [panel~(a2)] by a dashed-dotted brown line. For sufficiently large values of  $\alpha_{\mathrm{loc}}(0)$, we find that the estimated time $\tau_m$ indicates fairly well the delocalization time found from the numerical simulations, \ie\ the time where the results corresponding to the correlated disorder start separating significantly from those corresponding to the white-noise disorder. For the smallest value of $\alpha_{\mathrm{loc}}(0)$ [panel~(a1)], however, the estimate is rather poor. This is due to the fact that the edge of the wave packet at $\xi_0(\tau)$ is very smooth for low values of $\alpha$ [see Fig.~\ref{fig:density}(a) for white-noise disorder], although we found that it becomes slightly sharper for correlated disorder compared to white-noise disorder. This effect enhances the increase of the center-of-mass position for correlated disorder as compared to white-noise disorder. Hence the two curves separate significantly before the expected time $\tau_1$.

Consider now the cases where the moment $\overline{\xi^m}$ already diverges in the white-noise disordered potential, \ie\ $\alpha_{\mathrm{loc}}(0)>\alpha_m$.
In this case, we may intuitively guess that the behaviors of $\overline{\xi^m}$ in the white-noise and correlated disordered potentials start to differ significantly when the relative change of $\alpha(\tau)$ with respect to its initial value is of order 1. It corresponds to the time $\tau^*$ such that 
\begin{equation}
\frac{\Delta \alpha(\tau^*)}{\alpha_{\mathrm{loc}}(0)}=\frac{\alpha(\tau^*)-\alpha_{\mathrm{loc}}(0)}{\alpha_{\mathrm{loc}}(0)}=1.
\end{equation}
The corresponding times are shown as vertical orange lines on Figs.~\ref{fig:graph_x_avr_corr}(b1), (c1), and (c2).
We indeed find that they indicate fairly well the time when the behaviors in the white-noise and correlated disordered potentials start to separate significantly. This confirms that the dynamics of the position moments is mostly governed by the expansion of the edge of the wave packet and the disorder locally experienced by the particles located at this edge.

In both cases, the argument for delocalization is based on the property that the quantity $\alpha(\tau)$ increases in time. Since $k(x)$ always increases with $x$ for a finite force and $\xi_0(\tau)$ increases in time, both without upper bound, it is sufficient that the disorder power spectrum $\tilde{C}(2k)$ is a decreasing function.
Except for white-noise disorder where $\tilde{C}(2k)$ is a constant, this condition is almost always fulfilled~\cite{exceptions}.

\section{Conclusion}
\label{sec:conclusion}
In summary, we have studied the expansion dynamics of a quantum wave packet in a one-dimensional disordered potential in the presence of a constant bias force.
When the initial confinement is released, we found that the wave packet expands asymmetrically owing to the force dragging in one direction.
For white-noise disorder, the density profile progressively acquires power-law decaying tails, $n(x) \sim 1/x^{\eta}$. In the direction of the force, where the wave packet expands preferentially, we find that the exponent $\eta$ decreases smoothly as a function of the ratio of the force to the disorder strength $\alpha$. We found no evidence of any critical behavior on this quantity for any value of $\alpha$.
Nevertheless, algebraic localization is characterized by a series of critical values $\alpha_m$, where the $m$-th position moment diverges.
For $\alpha<\alpha_m$, we found that the $m$-th moment converges to a finite value compatible with the predictions of infinite-time diagrammatic calculations.
For $\alpha>\alpha_m$ we found that the $m$-th moment increases as a power law,
$\overline{x^m} \sim t^{\beta_m}$. Both $\beta_1$ and $\beta_2$ increase from $\beta_m =0$ for $\alpha=\alpha_m$ to $\beta_m \simeq 2$ for $\alpha \rightarrow +\infty$.
For correlated disorder, we found systematic delocalization of the expanding wave packet, irrespective to the model of disorder or correlation length as long as it is finite. More precisely, we identify a two-step dynamics, where both the center-of-mass position and the width of the wave packet show transient localization, similar to the white-noise case, and then delocalization at sufficiently long time. This correlation-induced delocalization is interpreted as due to the decrease of the effective de Broglie wave length, which lowers the effective strength of the disorder in the presence of finite-range correlations.

The expansion scheme we have considered in this work is the standard one proposed in Refs.~\cite{lsp2005,lsp2007,*lsp2007erratum,lsp2008} and used experimentally to demonstrate Anderson localization of ultracold matter waves~\cite{billy2008,roati2008,kondov2011,jendrzejewski2012}.
The additional, uniform bias field, whose effect on localization is studied here, may be implemented in such experiments using the gravity or a magnetic field gradient. The bias force may be controlled by combining the two in opposite directions and imposing a controlled imbalance between the two or controlling the inclination of the guide confining the atoms in 1D geometry.
The disordered potential may be realized in various ways, including
speckle light fields~\cite{clement2006,shapiro2012},
impurity atoms~\cite{gavish2005,paredes2005,gadway2011},
and shaped light fields controlled by digital mirror devices~\cite{choi2016} for instance. 
Algebraic localization in the presence of a bias force field may then be observed similarly as in previous ultracold-atom experiments by directly imaging the atomic density after a sufficiently long expansion time.
Saturation of the $m$-th position moment for $\alpha<\alpha_m$ and power-law expansion for $\alpha>\alpha_m$ may be observed by imaging the cloud at a variable time and computing the moments from the density profile.
This physics that is specific to white-noise disorder occurs at finite distances below the crossover value where the delocalization effects due to the finite correlations appear.
In the case of speckle-light disorder, a stronger delocalization effect is expected due to the well-known high-momentum cut-off of the disorder spectrum~\cite{lsp2007,*lsp2007erratum,gurevich2009,lugan2009}.

\section*{Acknowledgments}
We are grateful to Boris Altshuler, Gilles Montambaux, and Marie Piraud for useful discussions.
We thank Guilhem Bo\'eris for providing us the numerical routine to generate the disordered potentials.
This research was supported by the
European Commission FET-Proactive QUIC (H2020 grant No. 641122) and the European Research Council grant ERC-319286-QMAC.
It was performed using HPC resources from GENCI-CCRT/CINES (Grant c2015056853).
Use of the computing facility cluster GMPCS of the LUMAT federation (FR LUMAT 2764) is also acknowledged. 
The Flatiron Institute is supported by the Simons Foundation.
The content of this paper does not reflect the official opinion of the European Union. Responsibility for the information and views expressed therein lies entirely with the authors.

\begin{appendix}  

\section{DIAGRAMMATIC SOLUTION FOR THE PROBABILITY OF TRANSFER IN WHITE-NOISE DISORDER}
\label{sec:diagsol}
For non-correlated 1D Gaussian disorder, the probability of transfer can be calculated exactly using the diagrammatic approach of Ref.~\cite{berezinskii1974}. The latter has been extended to white-noise disorder in the presence of a bias force in Ref.~\cite{prigodin1980}.
For high particle energy, $\varepsilon\ll 1$, \ie, $E \gg (\hbar^2 F^2/2m)^{1/3}$, the probability of transfer reads as
\begin{widetext}
\begin{equation}
\label{eq:p_Q_R}
p(\xi,\tau|\alpha,\varepsilon)=\frac{1}{8\pi^2}\int \dd \omega\ \frac{\e^{-i\omega\tau}}{\sqrt{1+\xi}}
\ \int_0^{2\pi}\dd\varphi\ Q(\e^{-i\varphi},s)\Big[1+(1+\e^{i\varphi})R(\e^{i\varphi},s)\Big],
\end{equation}
where
\begin{equation}
s(\xi) = \frac{1}{2\alpha}\ln(1+\xi),
\label{eq:s_value}
\end{equation}
is the emerging dimensionless metrics in a bias disorder material~\cite{ccdb2017a}
and where the functions $Q(r,s)$ and $R(\rho,s)$ are the regular solutions of the two independent equations
\begin{equation}
\label{eq:diff_Q}
\frac{\partial Q}{\partial s}=\left[i\nu(s,\omega)-2\right] r\frac{\partial Q}{\partial r}+\left[\frac{i\nu(s,\omega)}{2}-1\right]Q - 2r \frac{\partial}{\partial r}\left(r\frac{\partial Q}{\partial r}\right)+
r\frac{\partial}{\partial r}\left(r\frac{\partial rQ}{\partial r}\right)+\frac{\partial}{\partial r}\left(r\frac{\partial Q}{\partial r}\right)
\end{equation}
and
\begin{equation}
\label{eq:diff_R}
\frac{\partial R}{\partial s} = -1+ \Big[-i\nu(s,\omega)+2-4\rho\Big] R +
\Big[{-i\nu(s,\omega)\rho}+6\rho-3\rho^2-1\Big]\frac{\partial R}{\partial \rho} -
\rho(1-\rho)^2 \frac{\partial^2 R}{\partial \rho^2},
\end{equation}
\end{widetext}
with
\begin{eqnarray}
\nu(s,\omega \vert \alpha,\varepsilon)
& = &
\frac{4\alpha}{\varepsilon}(1+\xi)^{3/2}\left[ \left(1 + \frac{\omega \varepsilon}{1+\xi}  \right)^{3/2} - 1 \right.
\nonumber \\
& & \left. - \frac{\omega \varepsilon}{1+\xi} \left( 1 + \frac{\omega \varepsilon}{1+\xi} \right)^{1/2} \right].
\nonumber
\end{eqnarray}
The initial conditions of the differential equations above are
$R(\rho,-\infty)=\frac{1}{1-\rho}$
and $Q(r,0)=\frac{1+r}{2}L(r,0)+\frac{1}{2}$,
where $L(r,s)$ is the solution of a differential equation similar to Eq.~(\ref{eq:diff_R}) but with opposite sign with the same initial condition, $L(r,-\infty)=\frac{1}{1-r}$.

In the long time limit, $\omega \ll 1/\varepsilon$, one thus finds
\begin{equation}
\label{eq:nu}
\nu(s,\omega|\alpha,\varepsilon) \simeq 2\alpha\omega \e^{\alpha s} =  2\alpha\omega \sqrt{1+\xi}.
\end{equation}
Hence, the dependence on the parameter $\varepsilon$ disappears. Since $\varepsilon$ does not appear explicitly in Eqs.~\eqref{eq:p_Q_R}-\eqref{eq:nu} either, for long propagation times, $\tau \gg \varepsilon$, \ie, $t\gg\hbar/E$,
the probability of transfer only depends on the parameter $\alpha$, that is on the ratio of the force to the disorder strength, $p(\xi,\tau|\alpha)$.

\section{EXPONENTIAL LOCALIZATION FOR A VANISHINGLY SMALL FORCE}
\label{sec:Andersonlimit}
For a vanishingly weak force, $F \rightarrow 0$, the length scale $\ell_0$ diverges and it is worth turning back to dimensionfull quantities.
More precisely, the limit $F \rightarrow 0$
for a fixed energy $E$, a fixed disorder strength $\Ur$, and a fixed distance $x$ yields
$\alpha \rightarrow 0_+$ [see Eq.~(\ref{eq:epsilonANDalpha})],
$\ell_0 \rightarrow +\infty$ [see Eq.~(\ref{eq:ellANDtau})],
and $\eta_\pm \simeq \pm 1/8\alpha$ [see Eq.~(\ref{eq:eta_pm})]
with $L_\textrm{\tiny loc} \equiv 4 \ell_0\alpha = 4\hbar^2 E / m\Ur$. The latter is the localization length in the absence of a force~\cite{lifshits1988}. Hence,  one finds
$\xi=x/\ell_0 \rightarrow 0_+$,
$f(\alpha \rightarrow 0,\lambda) \simeq \lambda \sin \left(\pi\lambda\right) \left[\frac{1+\lambda^2}{1+\cosh\left(\pi\lambda\right)}\right]^2$ [see Eq.~(\ref{eq:falpha})],
$(1+\xi)^{\mp \lambda^2/8\alpha} \simeq \exp\left(-\lambda^2\vert x \vert/2L_\textrm{\tiny loc} \right)$,
and
\begin{eqnarray}
P (x \vert E) & = & \frac{\pi^2}{8L_\textrm{\tiny loc}} \int_0^\infty \dd\lambda\, \lambda \sinh(\lambda) \left[\frac{1+\lambda^2}{1+\cosh(\pi\lambda)}\right]^2 \\
&& \times \exp\left[-\frac{(1+\lambda^2)}{2L_\textrm{\tiny loc}} \vert x \vert \right],
\nonumber
\end{eqnarray}
which is equal to the probability of transfer calculated in the absence of a force~\cite{berezinskii1974,gogolin1976a,gogolin1976b} (see Refs.~\cite{lsp2007,*lsp2007erratum,piraud2011} where the same notations as here are used).
Up to algebraic corrections, this probability of transfer decreases exponentially in the large distance limit, $\vert x \vert \gg L_\textrm{\tiny loc}$, in both directions $x>0$ or $x<0$.

\section{POWER LAW EXPONENTS  OF THE FIRST TWO POSITION MOMENTS AND INFINITE-TIME EXTRAPOLATION}
\label{sec:appendixA}
In this appendix, we describe the extrapolation method used for finding the values of the exponents $\beta_1$ and $\beta_2$ at infinite times.

\begin{figure*}[t!]
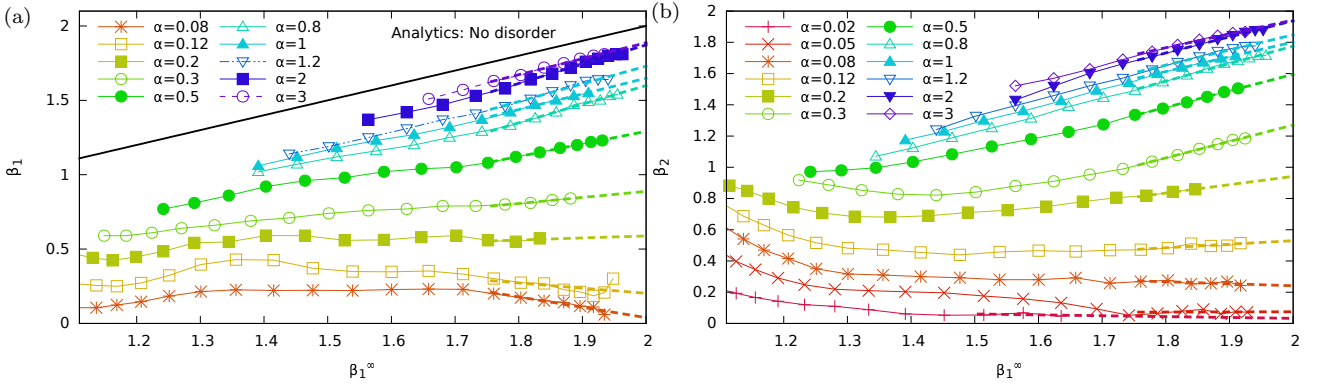

  \begin{tabular}{@{}p{0.48\linewidth}@{}p{0.48\linewidth}@{}}
    \subfigimg[width=\linewidth]{\footnotesize{\color{black} (a)}}{fig9a_beta1} &
    \subfigimg[width=\linewidth]{\footnotesize{\color{black} (b)}}{fig9b_beta2}
  \end{tabular}
\caption{\label{fig:power_laws}(Color online)
Power law exponent of
(a)~the average position
and
(b)~width of the wave packet as a function of $\beta_1^{\infty}$, for various values of the parameter $\alpha$.
The analytical result in the absence of disorder is shown as a black line in panel~(a).
Long-time linear extrapolations found from linear fits for $\beta_1^{\infty}>1.75$
are shown as dashed lines.
}
\end{figure*}

The smooth evolutions of $\overline{\xi}(\tau)$ and $\sigma_{\xi}(\tau)$ in log-log scale, visible on Fig.~\ref{fig:evolcumulants}, allow us to identify the local values of the power law exponents $\beta_1(\tau)$ and $\beta_2(\tau)$ as  the slopes of local linear fits of the curves.
In the absence of disorder, the exponent $\beta_1^{\infty}(\tau)=2(\tau+1)/(\tau+2)$ is a strictly increasing function of time, which evolves from $1$ at $\tau=0$ to $2$ at $\tau=2$.
Hence, $\beta_1^\infty(\tau)$ can serve as a measure of time, which advantageously converges to the finite value $\beta_1^{\infty}(\infty)=2$ in the infinite real time limit, $\tau \rightarrow \infty$.
On Fig.~\ref{fig:power_laws}, we plot the exponents $\beta_1(\tau)$ and $\beta_2(\tau)$ as functions of $\beta^\infty(\tau)$ for different values of $\alpha$. The quantity $\beta_1^{\infty}(\tau)$ found in the absence of disorder is also plotted on panel~(a).
We then extrapolate linearly the curves for $\beta_1$ and $\beta_2$ from the large time points $\beta_1^{\infty}>1.75$. It yields the estimates of $\beta_1(\tau=\infty))$ and $\beta_2(\tau=\infty))$ at $\beta_1^{\infty}=2$ plotted as black points on Figs.~\ref{fig:graph_beta_alpha}(a) and (b).

\begin{figure}
\includegraphics[width=0.49\textwidth]{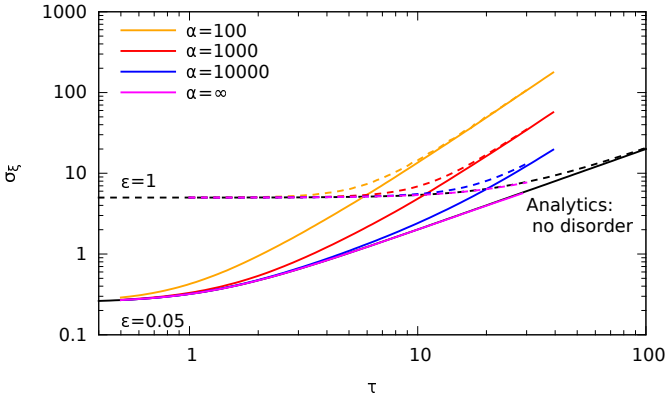}
\caption{\label{x_std_dev_2_forces}(Color online)
Time evolution of the width $\sigma_\xi$ of the wave packet as a function of time $\tau$ for two values of $\varepsilon$ and four values of the parameter $\alpha$.
The solid lines correspond to $\varepsilon=0.05$ and the dashed lines to $\varepsilon=1$.
The lower (upper) curves on the right-hand side correspond to the largest (smallest) values of $\alpha$ indicated on the panels.
The cases without disorder are plotted in dark black.
}
\end{figure}

For the sake of completeness, we have also studied the behavior of the width of the wave packet as a function of time for very large values of the parameter $\alpha$. Figure~\ref{x_std_dev_2_forces} shows that
for each value of $\varepsilon$, the width of the wave packet increases similarly as it does in the absence of disorder at short times.
At longer times, it then starts to increase faster and becomes independent of the value of $\varepsilon$.
The crossover time between the two regimes becomes larger when the values of $\alpha$ increase.

\section{PROPAGATION OF A GAUSSIAN WAVE PACKET SUBJECTED TO A BIAS FORCE}
\label{sec:appendixB}
Here, we study the evolution of the initial Gaussian wave packet
\begin{equation}
\psi(x,t=0)=\frac{1}{\pi^{1/4}\sqrt{\Delta x}}\e^{-x^2/2\Delta x^2+ip_0x/\hbar}
\end{equation}
in the presence of the uniform bias force $F$.
In momentum space, it reads as
\begin{equation}
\tilde{\psi}(p,t=0) =(2\pi)^{1/4}\sqrt{\frac{\hbar}{\sigma_p}}\e^{-(p-p_0)^2/4\sigma_p^2},
\end{equation}
with $p_0=\hbar k_0$ and $\sigma_p=\hbar/(\sqrt{2}\Delta x)$.
The eigenstate $\tilde{\psi}_E(p)$ of the Hamiltonian associated to the energy $E$ is the solution of the stationary Schr\"odinger equation
\begin{equation}
\frac{p^2}{2m} \tilde{\psi}_E(p)-iF\hbar\frac{\partial \tilde{\psi}_E(p)}{\partial p}=E \tilde{\psi}_E (p),
\end{equation}
the solution of which reads as
\begin{equation}
\tilde{\psi}_E(p)=\frac{1}{\sqrt{F}}\e^{\frac{i}{F\hbar}(Ep-p^3/6m)}.
\label{eq:phi_E}
\end{equation}
Then, decomposing the initial state on this eigenbasis, we find
\begin{equation}
\tilde{\psi}(p,t)=(2\pi)^{1/4}\sqrt{\frac{\hbar}{\sigma_p}}\e^{-\frac{(p-p_0-Ft)^2}{4\sigma_p^2}}\e^{-\frac{it[F^2t^2+3p(p-Ft)]}{6m\hbar}}.
\label{eq:phi_p_t}
\end{equation}
Applying inverse Fourier transformation to this solution, we then find
\begin{eqnarray}
\psi(x,t) &=& \int \frac{\dd p}{2\pi\hbar}\e^{ipx/\hbar}\tilde{\psi}(p,t)\\
&=& \frac{1}{\sqrt{2\sqrt{2\pi}A\sigma_p\hbar}}
\e^{-(Ft+p_0)^2/4\sigma_p^2+B^2/4A-iF^2t^3/6m\hbar},
\nonumber
\end{eqnarray}
with
\begin{equation}
A = \frac{1}{4\sigma_p^2}+\frac{it}{2m\hbar}
\qquad
\textrm{and}
\qquad
B =\frac{Ft+p_0}{2\sigma_p^2}+\frac{i}{\hbar}\left(x+\frac{Ft^2}{2m}\right).
\end{equation}
It yields the density profile
\begin{eqnarray}
|\psi(x,t)|^2 &=& \frac{1}{2\sqrt{2\pi}\hbar\sigma_p|A|}\e^{-(Ft+p_0)^2/2\sigma_p^2}\e^{2\Re(B^2/4A)}
\nonumber \\
&=& \frac{1}{\sqrt{2\pi}\sigma_x(t)}\e^{-[x-{x_{0}(t)}]^2/2\sigma_{x,0}^2(t)}
\end{eqnarray}
with
\begin{equation}
{x_{0}(t)} =\frac{p_0t}{m}+\frac{Ft^2}{2m}
\end{equation}
and 
\begin{equation}
\sigma_{x,0}(t) =\frac{\hbar}{2\sigma_p}\sqrt{1+\frac{4t^2\sigma_p^4}{m^2\hbar^2}}.
\end{equation}
Hence, the wave packet remains Gaussian and spreads as in the absence of a force.
The dynamics of the center of mass is that of the classical particle with the same initial position and velocity.

In dimensionless units, the wave packets reads as
\begin{equation}
\phi(\xi,\tau)=\frac{1}{\sqrt{2\pi}\sigma_{\xi}(\tau)}\e^{-[x-{{\xi}_{cl}(\tau)}]^2/2\sigma_{\xi}^2(\tau)}
\end{equation}
with
\begin{equation}
{\xi_{0}(\tau)}=\tau^2+2\tau
\end{equation}
and 
\begin{equation}
\sigma_{\xi,0}(\tau)
=\frac{\varepsilon}{\sqrt{2}}\frac{\kappa_0}{\Delta\kappa}\sqrt{1+4\frac{\tau^2}{\varepsilon^2}\left(\frac{\Delta\kappa}{\kappa_0}\right)^4}.
\end{equation}

\end{appendix}


\begin{thebibliography}{10}
\bibitem{corrdis}
{In the case of a correlated disorder, there is one additional parameter,
  namely the correlation length $\sigmar$, see
  Sec.~\ref{sec:correlated_disorder}}.
\bibitem{mthmomentfinte}
{Note that the $m$-th moment is finite also for $\eta_+ = m+1$ due to
  logarithmic corrections to the leading decay in $1/x^{m+1}$}.
\bibitem{converge}
{Note that the point at $\alpha \simeq 0.1$ has not yet reached the
  infinite-time limit as can be seen on Fig.~\ref{fig:evolcumulants}. Its
  asymptotic value is thus expected to be larger than the value found for the
  maximum time we have been able to compute}.
\end{thebibliography}
\end{document}